\documentclass[preprint,preprintnumbers,nofootinbib,amsmath,amssymb]{revtex4-1}
\usepackage{amssymb,latexsym,amsmath}

\usepackage[activeacute,english]{babel}
\usepackage{fancyhdr}
\usepackage{textcomp}
\usepackage{amsfonts}
\usepackage{graphicx}
\usepackage{graphics}
\usepackage{subfig}

\usepackage{xcolor}
\newtheorem{Theorem}{Theorem}
\newcommand{\pa}{\partial}

\newcommand{\si}{\sigma}

\newcommand{\om}{\omega}
\newcommand{\Om}{\Omega}

\newcommand{\rar}{\rightarrow}

\def\be {\begin {equation}}
\def\ee {\end {equation}}

\newcommand{\ba}{\begin{array}}
\newcommand{\ea}{\end{array}}
\newcommand{\bea}{\begin{eqnarray}}
\newcommand{\eea}{\end{eqnarray}}
\newcommand{\bi}{\begin{itemize}}
\newcommand{\ei}{\end{itemize}}

\begin{document}

\title{Classical $n$-body system in geometrical and volume variables. I. Three-body case}

\author{A. M. Escobar-Ruiz}
\email{admau@xanum.uam.mx}
\affiliation{
Departamento de F\'isica,
Universidad Aut\'onoma Metropolitana-Iztapalapa, San Rafael Atlixco 186,
M\'exico, CDMX, 09340 M\'exico}

\author{R. Linares}
\email{lirr@xanum.uam.mx}
\affiliation{
Departamento de F\'isica,
Universidad Aut\'onoma Metropolitana-Iztapalapa, San Rafael Atlixco 186,
M\'exico, CDMX, 09340 M\'exico}

\author{Alexander V Turbiner}
\email{turbiner@nucleares.unam.mx}
\affiliation{Instituto de Ciencias Nucleares, UNAM, M\'exico DF 04510, Mexico}

\author{Willard Miller, Jr.}
\email{miller@ima.umn.edu}
\affiliation{School of Mathematics, University of Minnesota, Minneapolis MN 55455, U.S.A.}

\begin{abstract}

We consider the classical 3-body system with $d$ degrees of freedom $(d>1)$ at zero total angular momentum. The study is restricted to potentials $V$ that depend solely on relative (mutual) distances $r_{ij}=\mid {\bf r}_i - {\bf r}_j\mid$ between bodies. Following the proposal by Lagrange, in the center-of-mass frame we introduce the relative distances (complemented by angles) as generalized coordinates and show that the kinetic energy does not depend on $d$, confirming results by Murnaghan (1936) at $d=2$ and van Kampen-Wintner (1937) at $d=3$, where it corresponds to a 3D solid body. Realizing $\mathbb{Z}_2$-symmetry $(r \rar -r)$ we introduce new variables $\rho=r^2$, which allows us to make the tensor of inertia non-singular for binary collisions. The 3 body positions form a triangle (of interaction) and the kinetic energy is $\mathcal{S}_3$-permutationally invariant wrt interchange of body positions and masses (as well as wrt interchange of edges of the triangle and masses). For equal masses, we use lowest order symmetric polynomial invariants of $\mathbb{Z}_2^{\otimes3} \oplus \mathcal{S}_3$ to define new generalized coordinates, they are called the {\it geometrical variables}. Two of them of the lowest order (sum of squares of sides of triangle and square of the area) are called {\it volume variables}. It is shown that for potentials, which depend on geometrical variables only (i) and those which depend on mass-dependent volume variables alone (ii), the Hamilton's equations of motion can be considered as being relatively simple. In the case (ii) all trajectories are mass-independent!

We study three examples in some detail: (I) 3-body Newton gravity in $d=3$, (II) 3-body choreography on the algebraic lemniscate by Fujiwara et al, and (III) the (an)harmonic oscillator.

\end{abstract}

%Use showkeys class option if keyword
                              %display desired

\maketitle

\section{Introduction}
\label{Introduction}

The Lagrangian for a classical system of two non-relativistic particles with masses $m_1$ and $m_2$, each of them with $d$ degrees of freedom, has the form
\begin{equation}\label{H2bo}
  {\cal L}_2 \ = \ \frac{1}{2}m_1\, {\dot{{\bf r}}_1}^{2}\ + \ \frac{1}{2}m_2\, {\dot{{\bf r}}_2}^{2} \ - \ V(\mid {\bf r}_1-{\bf r}_2\mid ) \ ,
\end{equation}
where ${\bf r}_i\in \mathbb{R}^d$ denotes the position vector of the $i$th particle, $\dot{{\bf r}}_i \equiv \frac{d}{dt}{\bf r}_i $ its velocity and $V$ is a translational invariant interaction potential with rotational symmetry. Solving this well known system is one of the basic problems in classical mechanics: it appears in many textbooks. After separation of the $d$-dimensional center-of-mass (cms) motion, the $2d$-dimensional problem is reduced to a $d$-dimensional one in the \emph{space of relative motion}. In the particular case of \emph{zero total angular momentum} the number of degrees of freedom in the Lagrangian (\ref{H2bo}) is reduced to one. Thus, the relative motion is described by the Lagrangian
\begin{equation}\label{H2bored}
  \tilde {\cal L}_2 \ = \ \frac{1}{2}\mu\, {\dot r}^{2}  \ - \ V(r) \ ,
\end{equation}
where $\mu=\frac{m_1\,m_2}{m_1+m_2}$ is the reduced mass of the system. The \emph{dynamical variable} $r\equiv \mid {\bf r}_1-{\bf r}_2\mid$ has an elementary \emph{geometrical interpretation}: it is the length of the interval which connects two particles. We call this the \emph{interval of interaction}.

\bigskip

In the case of three bodies with $d$-degrees of freedom, $d\geq1$, the number of relative distances is equal to the number of edges of the triangle which is formed by taking the bodies positions as vertices. This triangle defines the natural geometrical structure in which the interaction of the bodies occurs. We call it the {\it triangle of interaction}. In 1936, using the 3 mutual distances $r_{ij}$ (edges of the triangle of interaction) as generalized coordinates ($r-$representation), Murnaghan \cite{Murnaghan} introduced a canonical transformation to reduce the planar problem ($d=2$) from the configuration space $\mathbb{R}^6$ to $\mathbb{R}^3$. It implies that the original configuration space $\mathbb{R}^6$ is decomposed into the product $\mathbb{R}^2 \times \mathbb{R}_{\rm rel}^3 \times \mathcal{S}^1$ where the first factor corresponds to the planar cms motion whilst the last one corresponds to the cyclic (angular) motion of the system around the center-of-mass. Therefore, the problem is formulated in the three-dimensional space $\mathbb{R}_{\rm rel}^3 \subset \mathbb{R}^3$ of the relative motion. The natural extension of this reduction to the case of $3$-degrees of freedom ($d=3$) was then elaborated in Ref.\cite{Kampen}.

It follows from Ref.\cite{Murnaghan} and Ref.\cite{Kampen} that at zero total angular momentum, for both two and three degrees of freedom ($d=2,3$) the space of relative motion is the same $\mathbb{R}_{\rm rel}^3$ and the corresponding reduced Hamiltonians defined in a six-dimensional phase space $\mathbb{R}_{\rm rel}^3\times\mathbb{R}^3$ coincide. It will be shown in this paper that it remains true for any number of degrees of freedom, $d>1$. Let us denote this reduced Hamiltonian as ${\cal H}_{0}$. It has the form of the
kinetic energy of the 3D solid body with a certain tensor of inertia plus external potential.
Following identification of the coefficients of the tensor of inertia in ${\cal H}_{0}$ as the entries of a contravariant metric (cometric), the emerging Hamiltonian describes a three-dimensional particle moving in a curved space with cometric $g^{\mu\,\nu}(r_{ij})$. Remarkably, the components of $g^{\mu\,\nu}(r_{ij})$ are rational functions of $r_{ij}$ as well as its determinant\footnote{This result was known by Murnaghan \cite{Murnaghan}.}.

\bigskip

The triangle of interaction is characterized by three edges $r_{ij}$ (intervals of interaction). It is easy to see that the free Hamiltonian ${\cal H}_{0}$, i.e. when the potential $V$ is absent, is $\mathbb{Z}_2^{\otimes 3}$-invariant under the reflections $r_{ij}\rightarrow -r_{ij}$ \cite{Murnaghan}. It seems natural to make symmetry reduction introducing $\mathbb{Z}_2$-invariant coordinates $\rho_{ij}=r_{ij}^2$ with their corresponding canonical momentum (it will be called $\rho-$representation). Lagrange in Ref.\cite{Lagrange} already paid attention to the importance of $r-$ and $\rho$-variables, but never used them as dynamical coordinates (generalized coordinates). The corresponding symmetric reduced Hamiltonian was never constructed by him explicitly. It was done much later by Murnaghan\cite{Murnaghan} in $r-$variables only. Up to our present knowledge nobody used the $\rho$- or volume variables as dynamical variables (generalized coordinates). It is worth mentioning that Lama\^{\i}tre, in Ref.\cite{Lemaitre}, derived a reduced Hamiltonian using certain radial variables but those variables suffer lacking the property of $\mathcal{S}_3$ permutation invariance. In $\rho$-coordinates, we will see that the free Hamiltonian ${\cal H}_{0}$ is a second degree polynomial in momentum variables with linear $\rho$-dependent coefficients on the phase space \cite{TME3-d}. In the case of equal masses the free Hamiltonian ${\cal H}_{0}$ has extra permutational symmetry $\mathcal{S}_3$ with respect to permutations of the intervals of interaction $r_{ij}$ (or their squares $\rho_{ij}$).
It suggests the introduction of new $\mathcal{S}_3$-permutationally invariant coordinates in $\rho$-space
\begin{equation}
\label{sigv}
\begin{aligned}
 & \si_1 \ = \ \rho_{12} \  + \ \rho_{23} \  + \ \rho_{31} \  ,
\\
 & \si_2 \ = \ \rho_{12}\,\rho_{23} \  + \ \rho_{12}\,\rho_{31} \ + \ \rho_{23}\,\rho_{31} \  ,
\\
 & \si_3 \ = \ \rho_{12} \, \rho_{23} \, \rho_{31} \  .
\end{aligned}
\end{equation}
These coordinates, as well as ones of the original 3-body system, are invariant under the permutations of the bodies. In variables (\ref{sigv}), the free Hamiltonian ${\cal H}_{0}$ remains a polynomial in phase space $(\si, p_{\si})$, in agreement with Ref.\cite{TME3-d}. Following the Cayley-Menger formula \cite{CayleyMenger}, instead of $\si_2$ one can introduce the \emph{square} of the area of the triangle of interaction, $S = \frac{1}{16}(4\,\si_2 - \si_1^2)$. Making a canonical transformation one can show that the free Hamiltonian ${\cal H}_{0}$ in coordinates $(P \equiv \frac{1}{2}\si_1,\, S,\, T \equiv  \si_3)$ remains a polynomial. The variables $(P,S)$ are called the {\it volume variables}. The key task of this article is to derive the explicit form of the free Hamiltonian ${\cal H}_{0}$ that answers the question: \emph{What is the form of the reduced Hamiltonian in variables $(P,S,T)$?}

\bigskip

The goal of the present study is to write, for the 3-body system with zero total angular momentum and arbitrary $d>1$, the reduced Hamiltonian ${\cal H}_{0}$ in the $\rho-$representation and also in the $(P,S,T)$-representation. The motivation of the present paper is three-fold. Our first aim is to demonstrate that the choice of the intervals of interaction \emph{squared}, $\rho_{ij}=r^2_{ij}$, as dynamical variables leads to a deep connection between the geometrical characteristics of the triangle of interaction and the dynamics of the system. Accordingly, it will be shown that for arbitrary masses the corresponding Hamiltonian in $\rho$ variables describes a particle in a curved space with a certain (essentially non-flat) metric $g^{\mu\,\nu}(\rho)$. Unlike the $r-$representation, now the components of $g^{\mu\,\nu}(\rho)$ are first degree polynomials in $\rho$-variables and its determinant $\text{Det}[g^{\mu\,\nu}(\rho)]$ is proportional to the square of the area of the triangle of interaction.

Secondly, just as for the case of three identical particles $m_1=m_2=m_3=1$ we will show that the set of dynamical coordinates $(P,S,T)$ exhibit outstanding properties. In particular,
\begin{itemize}
  \item The Det$\big[g^{\mu\,\nu}(P,S,T)\big]$ is, in fact, the discriminant of the fourth degree polynomial equation that defines the physically relevant 3-body Newtonian gravity potential at $d=3$ in terms of the aforementioned variables.

  \item Remarkably, for planar 3-body choreographic motion on an algebraic lemniscate by Jacob Bernoulli, studied by Fujiwara et al.\cite{Fujiwara}, two of the variables $P,T$ become \emph{particular} constants of motion \cite{ATJC,Turbiner:2013p} and the trajectory is an elliptic curve.
\end{itemize}

\bigskip

Thirdly, we investigate how the volume variables $P,S$ are modified in the case of arbitrary masses. Assuming that the potential $V$ depends on above-introduced volume variables $(P,S)$, we will show that for the original Hamiltonian there exist trajectories which depend on these two volume variables alone.

\bigskip

The main long term goal for doing this work is to explore the possible practical advantages of representations for which the involved generalized coordinates posses the following two properties: (i) they encode the symmetries of the free reduced Hamiltonian ${\cal H}_{0}$, and (ii) ${\cal H}_{0}$ is a polynomial function in these variables. The reason to consider such variables comes not only from esthetics (being esthetics a sufficient reason, though). Upon the standard quantization procedure the associated Hamiltonian operator can be constructed from the classical one. Then, due to the property (ii) the kinetic energy term will be an algebraic operator, i.e. the coefficients in front of the derivatives are polynomial functions. Algebraic operators has been intimately related with the existence of exactly and quasi-exactly-solvable quantum models. Hence, we think that a study of classical systems within the aforementioned representations makes sense and it is worthwhile to pursue.

\bigskip

Also, in the general $n$-body case in $d-$dimensions ($n\geq d-1$) the dynamics of the system can be thought of as a type of "breathing" \emph{polytope of interaction}. At fixed $n$ there exist ($n-1$) \emph{volume-variables} made out of geometric elements with different dimensionality (edges, faces, cells and so on) of this polytope of interaction, each variable being $\mathcal{S}_n$-permutationally symmetric. Therefore, another rationale of the present consideration is to shed light on the links between the theory of regular polytopes, fundamental symmetric polynomials and the dynamics of an $n$-body system.

\bigskip

The structure of the article is organized very simply. In Section \ref{planar case}, we first review the symmetric reduction of the planar 3-body system. Separating the cms motion and using the conserved total angular-momentum $p_{\Omega}$, the problem is reduced to one of three degrees of freedom in which the coordinates are the lengths $r_{12},r_{23},r_{31}$ of the sides of the triangle of interaction. It is shown that the corresponding reduced Hamiltonian ${\cal H}_0$ describes a $3$-dimensional particle moving in a curved space. Next, we introduce the variables $\rho_{ij}=r^2_{ij}$ ($\rho$-representation) for which ${\cal H}_0$ becomes a polynomial at $p_{\Omega}=0$. In Section \ref{non planar case} we prove that for zero total angular momentum and arbitrary dimension $d>1$ the trajectories are described by exactly the same Hamiltonian ${\cal H}_0$. The $3$-body chain of harmonic oscillators is briefly revisited to illustrate this representation. The case of three identical particles $m_1=m_2=m_3=1$ is studied in Section \ref{PST}. For ${\cal H}_0$, the novel set of variables $(P,S,T)$ is introduced and the corresponding equations of motion are presented explicitly. The $3$-body Newtonian gravity potential in $d=3$ and planar choreographic trajectories on algebraic lemniscate ($d=2$) are used to exemplify the properties of this representation. Finally, in Section \ref{volumeR} in the case of arbitrary masses with a certain class of potentials a further \emph{reduction} of the problem to one of two degrees of freedom is accomplished using modified, mass-dependent volume variables as dynamical coordinates. For conclusions and future
outlook see Section \ref{Conclusion}.

\bigskip

\section{Three-body system: planar case ($d=2$)}
\label{planar case}

\bigskip

We consider a planar ($d=2$) classical system of three interacting point-like particles with masses $m_1,m_2$ and $m_3$, respectively. The Lagrangian is of the form,
\begin{equation}
\label{H}
   {\cal L}\ =\ {\cal T} \ - \ V(r_{12},\,r_{23},\,r_{31})  \ ,
\end{equation}
where the kinetic energy is given by
\begin{equation}
\label{Tflat}
   {\cal T}\ =\ \frac{1}{2}m_1\, {\dot{{\bf r}}_1}^{2}\ + \ \frac{1}{2}m_2\, {\dot{{\bf r}}_2}^{2}\ + \ \frac{1}{2}m_3\, {\dot{{\bf r}}_3}^{2}\ ,
\end{equation}
here ${\bf r}_i \in \mathbb{R}^2$ is the vector position of the $i$th body, $\dot{{\bf r}_i} \equiv \frac{d}{dt}{\bf r}_i $ and the potential $V$ depends on the relative distances
\[
r_{ij} \ \equiv \  \mid {\bf r}_i-{\bf r}_j\mid \ ,
\]
between particles only. This implies that ${\cal L}$ is rotationally symmetric. The configuration space is six-dimensional $\mathbb{R}^2({\bf r}_1) \times \mathbb{R}^2({\bf r}_2) \times\mathbb{R}^2({\bf r}_3)$.

The kinetic energy (\ref{Tflat}) can be expressed in terms of center-of-mass and the relative coordinates of the three bodies \cite{Murnaghan},
\begin{equation}
\label{Tflat2}
\begin{aligned}
   {\cal T}\ & = \ \frac{1}{2}M\,{\dot{\bf Y}}^2 \ + \  \frac{1}{2}{\mu_{12}} \,{\dot{{\bf r}}_{12}}^{2}\ + \ \frac{1}{2}{\mu_{23}}\, {\dot{{\bf r}}_{23}}^{2}\ + \ \frac{1}{2}{\mu_{31}}\, {\dot{{\bf r}}_{31}}^{2}
\\ &
  = \ \frac{1}{2}M\,{\dot{\bf Y}}^2 \ + \ \frac{1}{2}{\mu_{12}}\,( {\dot r}_{12}^2 \ + \ {r}_{12}^2 \,{\dot \theta}_{12}^2  ) \ + \ \frac{1}{2}{\mu_{23}}\,( {\dot r}_{23}^2 \ + \ {r}_{23}^2 \,{\dot \theta}_{23}^2  ) \ + \ \frac{1}{2}{\mu_{31}}\,( {\dot r}_{31}^2 \ + \ {r}_{31}^2 \,{\dot \theta}_{31}^2  )\ ,
\end{aligned}
\end{equation}
where $M=m_1+m_2+m_3$ is the total mass, ${\bf Y} =  \frac{m_1\,{\bf r}_1  +  m_2\,{\bf r}_2  +  m_3\,{\bf r}_3}{M}$ is the center-of-mass vector, and $\mu_{ij} \equiv \frac{m_i\,m_j}{M}$ denotes a reduced-like mass. In (\ref{Tflat2}), we also introduce polar coordinates for relative vectors, i.e. ${\bf r}_{ij}\equiv (r_{ij},\,\theta_{ij})$ in the space of relative motion.

For future convenience, we select the center-of-mass system as the inertial frame, ${\bf Y} =0, {\dot{\bf Y}}=0$. Eventually, in this inertial frame the system is characterized by four degrees of freedom.

\subsection{$r$-representation}

Now, following Ref.\cite{Murnaghan} we consider the three \emph{relative distances} $r_{12},\,r_{23},\,r_{31},\,$ and the \emph{angle} \begin{equation}\label{AngOm}
  3\,\Omega \ = \ \theta_{12}\ + \ \theta_{23} \ + \ \theta_{31} \ ,
\end{equation}
as the four generalized coordinates of the Lagrangian (\ref{H}). It turns out that the kinetic energy (\ref{Tflat2}) does not depend on $\Omega$ (\ref{AngOm}). Therefore, the angle $\Omega$ is a cyclic variable and its canonical momentum
\begin{equation}\label{Pomega}
\begin{aligned}
  p_{{}_\Om}  \ & \equiv \ \frac{\pa}{\pa \dot \Om} \,{\cal L} \  = \ \frac{\pa}{\pa \dot \Om} \,{\cal T}
  \\ &
 = \  \mu_{12}\,{r}_{12}^2 \,{\dot \theta}_{12}^2 \ + \ \mu_{23}\,{r}_{23}^2 \,{\dot \theta}_{23}^2\ + \ \mu_{31}\,{r}_{31}^2 \,{\dot \theta}_{31}^2 \ ,
\end{aligned}
\end{equation}
is a constant of motion (saying differently, the first integral), $\dot p_{{}_\Om}=0$. From (\ref{Pomega}) it follows that $p_{{}_\Omega}$ is nothing but the total angular momentum of the system about its center of mass. The conserved quantity $p_{{}_\Om}$ allows us to reduce the problem to one of three degrees of freedom in which the corresponding coordinates are the relative distances $r_{12},\,r_{23}$ and $r_{31}\,$ \cite{Murnaghan}. To this end, it is convenient to introduce the Routhian ${\cal R}$ defined \cite{LandauLifshitzBook} by the Legendre transformation
\begin{equation}
\label{Routhi}
{\cal R}(r_{ij},{\dot r}_{ij},p_{{}_\Om})\ = \ {\cal L}\ - \ p_{{}_\Om}\,{\dot \Om}    \ .
\end{equation}
It is a function of the radial variables $r_{ij},{\dot r}_{ij}$ and the integral $p_{{}_\Om}$ only.

\subsubsection{Hamiltonian for the reduced three-dimensional problem}

\begin{figure}[htp]
  \centering
  \includegraphics[width=13.0cm]{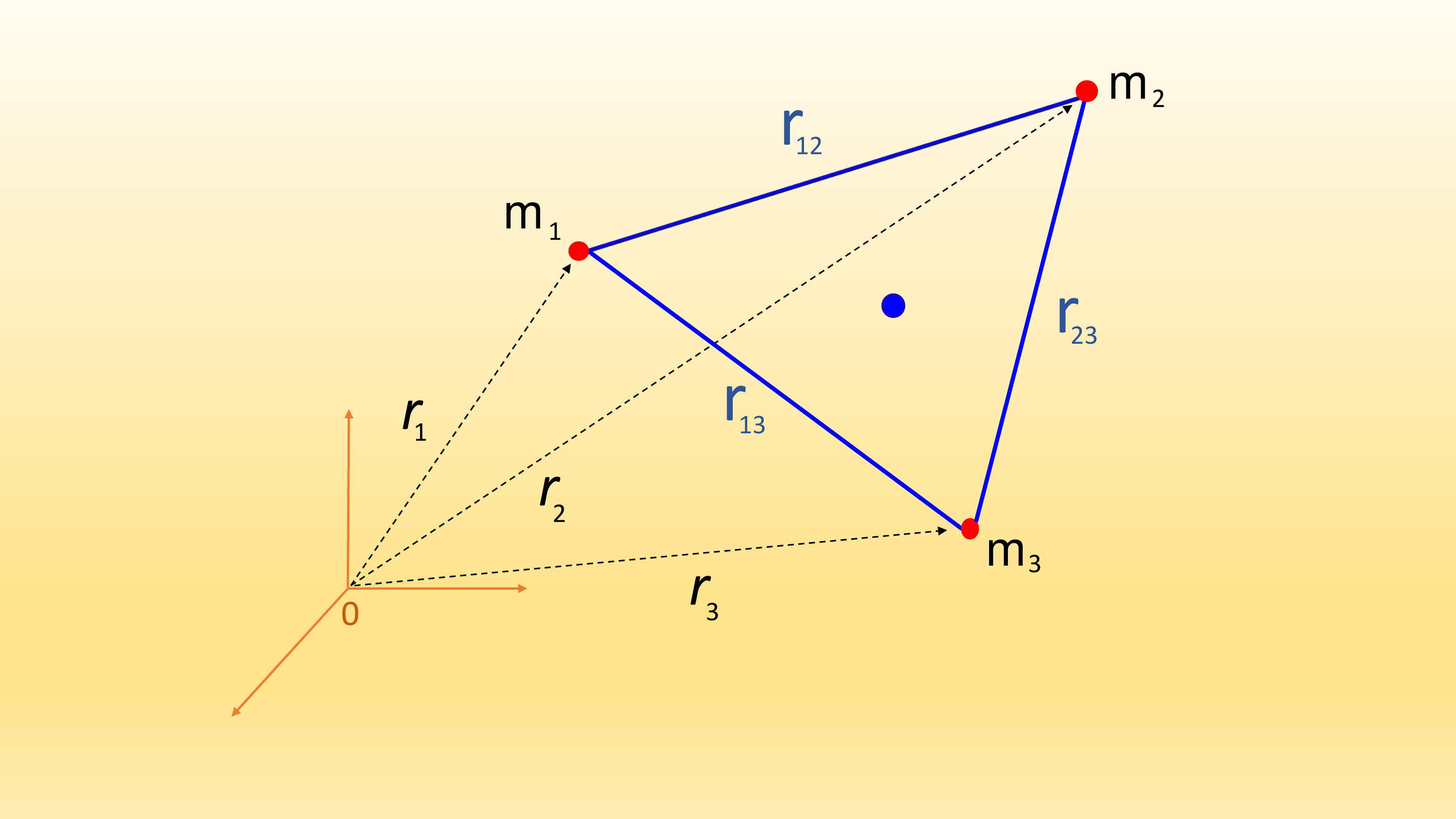}
  \caption{Triangle of interaction in $d=3$: the individual coordinate vectors ${\bf r}_i$ mark positions of vertices of the triangle with sides $r_{ij}$. The center-of-mass (the barycenter of the triangle) is marked by a (blue) bubble.}
\label{Fig3}
\end{figure}

In this Section we switch, for convenience, from the Lagrangian formalism to the Hamiltonian one. From (\ref{Routhi}) it follows that the reduced Hamiltonian of the system is defined in a six-dimensional phase space where the dynamical variables are the relative distances $r_{ij}$ and its conjugate momentum variables

\[
p_{12}\ \equiv \  \frac{\partial\,{\cal R}}{\partial {\dot r}_{12} }\ ,\qquad p_{23}\ \equiv \ \frac{\partial\,{\cal R}}{\partial {\dot r}_{23} }\ ,\qquad p_{31} \ \equiv \ \frac{\partial\,{\cal R}}{\partial {\dot r}_{31} } \ .
\]
In particular, for zero total angular momentum\footnote{In atomic and subatomic quantum systems the lowest energy state is usually one of zero total angular momentum ($S-$states)} $p_{{}_\Omega}=0$, the momentum variable $p_{12}$ reads
\begin{equation}
\label{p12g}
\begin{aligned}
&  p_{12} \ = \  \frac{1}{16\,I\, S^2_{\bigtriangleup}}\, \bigg[ \, 4 \,r_{12}^2\,\big [   \left(\mu _{12}\, \mu _{23}+\mu _{12}\, \mu _{31}+\mu _{23}\, \mu _{31} \right) r_{23}^2 \, r_{31}^2\ + \ 4 \,\mu _{12}^2 \,S^2\,  \big]\,{\dot r}_{12} \ - \
\\ &
r_{12}\, r_{23} \,\big[ \,2\, \left(\mu _{12}+\mu _{23}\right)\, \mu _{31} \left(r_{12}^2+r_{23}^2-r_{31}^2\right)\, r_{31}^2+\mu _{12}\, \mu _{23}\, \left(\left(r_{12}^2-r_{23}^2\right){}^2-r_{31}^4\right)    \big]\,{\dot r}_{23}  \ - \
\\ &
r_{12}\, r_{31} \,\big[ \,2\, \left(\mu _{12}+\mu _{31}\right)\, \mu _{23} \left(r_{12}^2-r_{23}^2+r_{31}^2\right)\, r_{23}^2+\mu _{12}\, \mu _{31}\, \left(\left(r_{12}^2-r_{31}^2\right){}^2-r_{23}^4\right)    \big]\,{\dot r}_{31} \bigg] \ ,
\end{aligned}
\end{equation}
here
\begin{equation}
\label{VT}
  I \ \equiv \  \mu _{12}\,r_{12}^2 \ + \  \mu _{23}\,r_{23}^2 \ + \ \mu _{31}\,r_{31}^2 \ ,
\end{equation}
is the moment of inertia with respect to the center of mass, and
\begin{equation}\label{VS}
  S^2_{\bigtriangleup} \ \equiv \  \frac{1}{16}\big(\,  2\,r_{12}^2\,r_{23}^2 \ +\  2\,r_{12}^2\,r_{31}^2 \ +\  2\,r_{23}^2\,r_{31}^2 \ -\  r_{12}^4 \ -\  r_{23}^4 \ -\  r_{31}^4 \,\big) \ ,
\end{equation}
is the \emph{square} of the area of the triangle formed by the three particles. It was called the \emph{triangle of interaction}, see {Fig. \ref{Fig3}.} By a cyclic arrangement of the labels $(1,2,3)$ (\,i.e. $(2,3,1)$ or $(3,1,2)$\,) in (\ref{p12g}) we obtain the other two momenta $p_{23}$ and $p_{31}$. The two geometrical quantities $P_m\equiv \frac{m_1+m_2+m_3}{2\,m_1\,m_2\,m_3} \,I$ and $S_m\equiv \frac{3\,m_1\,m_2\,m_3}{m_1+m_2+m_3}S^2_{\bigtriangleup}$ will play an important role in the present study. They will be called {\it modified volume variables}.

It is worth mentioning that in Ref.\cite{Hsiang} the virtual motions of the \emph{triangle of interaction}, through which the
triangle changes its kinematic invariants such as size, shape and orientation (position) and velocity are the starting point for a different systematic geometric-based approach to the three-body problem. However, unlike us, in Ref.\cite{Hsiang} $\mathcal{S}_n$ permutationally-invariant variables are not considered as the fundamental dynamical generalized coordinates.

Eventually, we arrive to the reduced Hamiltonian \cite{Murnaghan}
\begin{equation}
\label{Hred}
\begin{aligned}
  {\cal H}_{r} \  & =  \ p_{12}\,{{\dot r}_{12}} \ + \ p_{31}\,{{\dot r}_{31}} \ + \ p_{23}\,{{\dot r}_{23}} \ - \ {\cal R}
\\ &
  =\ \frac{1}{2}\bigg[\,\frac{p_{12}^2}{m_{12}} \ + \ \frac{p_{23}^2}{m_{23}} \ + \ \frac{p_{31}^2}{m_{31}} \ + \ \frac{r_{12}^2+r_{31}^2-r_{23}^2}{m_1\,r_{12}\,r_{31}}\,p_{12}\,p_{31}\ +
  \ \frac{r_{12}^2+r_{23}^2-r_{31}^2}{m_2\,r_{12}\,r_{23}}\,p_{12}\,p_{23}\ + \
\\ &
\qquad \frac{r_{23}^2+r_{31}^2-r_{12}^2}{m_3\,r_{23}\,r_{31}}\,p_{23}\,p_{31} \,\bigg] \ + \ \frac{2}{3}\,p_{{}_\Om}\,S_{\bigtriangleup}\bigg[\, p_{12}\,\frac{m_1\,r_{31}^2-m_2\,r_{23}^2}{m_1\,m_2\,r_{12}\,r_{23}^2\,r_{31}^2} \ + \ p_{23}\,\frac{m_2\,r_{12}^2-m_3\,r_{31}^2}{m_2\,m_3\,r_{23}\,r_{12}^2\,r_{31}^2}
\\ &
  \quad\ + \ p_{31}\,\frac{m_3\,r_{23}^2-m_1\,r_{12}^2}{m_1\,m_3\,r_{31}\,r_{23}^2\,r_{31}^2}\,\bigg]\ +
  \ V_{\text{eff}} \quad ,\qquad m_{ij}\equiv \frac{m_i\,m_j}{m_i+m_j} \ ,
\end{aligned}
\end{equation}
where
\[
V_{\text{eff}} \ = \ V  \ + \ \frac{p^2_{{}_\Om}}{9}\bigg[  \frac{1}{m_{12}\,r_{12}^2} \, + \,  \frac{1}{m_{23}\,r_{23}^2} \, + \,  \frac{1}{m_{31}\,r_{31}^2} \,-\, \frac{r_{12}^2}{2\,m_3\,r_{23}^2\,r_{31}^2}  \,-\, \frac{r_{23}^2}{2\,m_1\,r_{12}^2\,r_{31}^2} \,-\, \frac{r_{31}^2}{2\,m_2\,r_{12}^2\,r_{23}^2}  \bigg] \ ,
\]
is an effective potential. The angular momentum $p_{{}_\Om}$ is a Liouville integral, it Poisson-commutes with the Hamiltonian (\ref{Hred}), $\{p_{{}_\Om},\, {\cal H}_{r}\}=0$. At zero angular momentum $p_{{}_\Om}=0$ the second term in kinetic energy vanishes as well as the second term in the effective potential $V_{\rm eff}$. Explicitly, at $p_{{}_\Om}=0$ the above Hamiltonian (\ref{Hred}) becomes
\begin{equation}
\label{Hredzeroam}
\begin{aligned}
  {H}_{r} \  & = \ \frac{1}{2}\bigg[\,\frac{p_{12}^2}{m_{12}} \ + \ \frac{p_{23}^2}{m_{23}} \ + \ \frac{p_{31}^2}{m_{31}} \ + \ \frac{r_{12}^2+r_{31}^2-r_{23}^2}{m_1\,r_{12}\,r_{31}}\,p_{12}\,p_{31}\ +
  \ \frac{r_{12}^2+r_{23}^2-r_{31}^2}{m_2\,r_{12}\,r_{23}}\,p_{12}\,p_{23}\ + \
\\ &
\qquad \frac{r_{23}^2+r_{31}^2-r_{12}^2}{m_3\,r_{23}\,r_{31}}\,p_{23}\,p_{31} \,\bigg] \ + \  V(r_{12},\,r_{23},\,r_{31})     \ .
\end{aligned}
\end{equation}

The way how the reduced Hamiltonian (\ref{Hredzeroam}) is written we call the $r-$representation.

\subsubsection{The metric $g^{\mu \nu}(r)$}

The associated cometric for (\ref{Hred}) defined by coefficients in front of the quadratic terms in variables $(p_{12},\,p_{31},\,p_{23})$ is given by
\begin{equation}
\label{gmn33-red}
 g^{\mu \nu}(r)\ =
        \left|
 \begin{array}{ccc}
 \frac{1}{2\,m_{12}} & \ \frac{r_{12}^2 +r_{31}^2-r_{23}^2 }{4\,m_1\,r_{12}\,r_{31}} & \ \frac{r_{12}^2 +r_{23}^2-r_{31}^2 }{4\,m_2\,r_{12}\,r_{23}} \\
            &                                   &                                   \\
 \frac{r_{12}^2 +r_{31}^2-r_{23}^2 }{4\,m_1\,r_{12}\,r_{31}} & \  \frac{1}{2\,m_{31}} & \ \frac{r_{31}^2 +r_{23}^2-r_{12}^2 }{4\,m_3\,r_{23}\,r_{31}} \\
            &  \                                  &                                   \\
 \frac{r_{12}^2 +r_{23}^2-r_{31}^2 }{4\,m_2\,r_{12}\,r_{23}} & \ \frac{r_{31}^2 +r_{23}^2-r_{12}^2 }{4\,m_3\,r_{23}\,r_{31}} & \frac{1}{2\,m_{23}}
 \end{array}
        \right| \ .
\end{equation}
Its determinant $\mathcal{D}_m = \text{Det}[g^{\mu \nu}(r)]$ possesses the remarkable factorization property
\[
\mathcal{D}_m \ = \ \frac{{(m_1+m_2+m_3)}^2}{2\,m_1^2\,m_2^2\,m_3^2}\, \frac{I\, \ S^2_{\triangle}}{r_{12}^2\,r_{23}^2\,r_{31}^2}\ ,
\]
thus, it is proportional to the moment of inertia $I$ (\ref{VT}) and the area (squared) $S^2_{\bigtriangleup}$ (\ref{VS}) of the triangle of interaction.\\
\textbf{Remark.} The above determinant $\mathcal{D}_m$ is a rational function in $r$-variables. It effectively depends on the three coordinates $I$, $S^2_{\bigtriangleup}$ and $T \equiv r_{12}^2\,r_{23}^2\,r_{31}^2$. Note that it vanishes, $\mathcal{D}_m=0$, iff the area of the triangle of interaction is equal to zero or a triple-body collision occurs. It is singular at the two-body collision point.

\subsection{$\rho$-representation}

The cometric (\ref{gmn33-red}) as well as the original kinetic energy (\ref{Tflat2}) are invariant formally under reflections $\mathbb{Z}_2\oplus \mathbb{Z}_2\oplus \mathbb{Z}_2$,
\[
r_{12}\, \rightarrow \, - r_{12} \ , \qquad r_{23}\, \rightarrow \, - r_{23} \ , \qquad r_{31}\, \rightarrow \, - r_{31} \ ,
\]
and w.r.t. $\mathcal{S}_3$-group action (permutations of the bodies). If we introduce new variables,
\begin{equation}\label{rhovar}
\rho_{12}\, = \,  r_{12}^2 \ , \qquad \rho_{23}\, = \,  r_{23}^2 \ , \qquad \rho_{31}\, = \,  r_{31}^2 \ ,
\end{equation}
with the corresponding canonical momenta
\begin{equation}\label{prhovar}
P_{12}\, = \,  \frac{1}{2\,r_{12}}\,p_{12} \ , \qquad P_{23}\, = \,  \frac{1}{2\,r_{23}}\,p_{23} \ , \qquad P_{31}\, = \,  \frac{1}{2\,r_{31}}\,p_{31} \ ,
\end{equation}
we immediately arrive at the ${\mathbb{Z}}_2$-symmetry reduced Hamiltonian
\begin{equation}\label{Hredrho}
\begin{aligned}
  {\cal H}_{\rho} \  & = \ 2\,\bigg[\,\frac{\rho_{12}\,P_{12}^2}{m_{12}} \ + \ \frac{\rho_{23}\,\,P_{23}^2}{m_{23}} \ + \ \frac{\rho_{31}\,\,P_{31}^2}{m_{31}} \ + \ \frac{\rho_{12} + \rho_{31}- \rho_{23}}{m_1}\,P_{12}\,P_{31}\ + \ \frac{\rho_{12} + \rho_{23}- \rho_{31}}{m_2}\,P_{12}\,P_{23}
\\ &
 + \  \frac{\rho_{23} + \rho_{31}- \rho_{12}}{m_3}\,P_{23}\,P_{31}\,\bigg]\ + \ \frac{4}{3}\,p_{{}_\Omega}\,S_{\bigtriangleup}\bigg[\, P_{12}\,\frac{m_1\,\rho_{31}-m_2\,\rho_{23}}{m_1\,m_2\,\rho_{23}\,\rho_{31}} \ + \ P_{23}\,\frac{m_2\,\rho_{12}-m_3\,\rho_{31}}{m_2\,m_3\,\rho_{12}\,\rho_{31}}
\\ &
    + \ P_{31}\,\frac{m_3\,\rho_{23}-m_1\,\rho_{12}}{m_1\,m_3\,\rho_{23}\,\rho_{31}}   \,\bigg]
  \ + \ V_{\text{eff}} \ ,
\end{aligned}
\end{equation}
where
\[
V_{\text{eff}} \ = \ V  \ + \ \frac{p^2_{{}_\Omega}}{9}\bigg[  \frac{1}{m_{12}\,\rho_{12}} \, + \,  \frac{1}{m_{23}\,\rho_{23}} \, + \,  \frac{1}{m_{31}\,\rho_{31}} \,-\, \frac{\rho_{12}}{2\,m_3\,\rho_{23}\,\rho_{31}}  \,-\, \frac{\rho_{23}}{2\,m_1\,\rho_{12}\,\rho_{31}} \,-\, \frac{\rho_{31}}{2\,m_2\,\rho_{12}\,\rho_{23}}  \bigg] \ .
\]
The Hamiltonian (\ref{Hredrho}) is written in what we call the $\rho-$representation (cf. (\ref{Hred})).

\subsubsection{The metric $g^{\mu \nu}(\rho)$}

The associated contravariant metric for (\ref{Hredrho}) defined by coefficients in front of the quadratic terms in momentum variables $(P_{12},\,P_{31},\,P_{23})$
\begin{equation}
\label{gmn33-rho}
 g^{\mu \nu}(\rho)\ = \left|
 \begin{array}{ccc}
 2\,\frac{\rho_{12}}{m_{12}} & \ \frac{\rho_{12} + \rho_{31} - \rho_{23}}{m_1} & \ \frac{\rho_{12} + \rho_{23} - \rho_{31}}{m_2} \\
            &                                   &                                   \\
 \frac{\rho_{12} + \rho_{31} - \rho_{23}}{m_1} & \  2\,\frac{\rho_{31}}{m_{31}} & \ \frac{\rho_{31} + \rho_{23} - \rho_{12}}{m_3} \\
            &  \                                  &                                   \\
 \frac{\rho_{12} + \rho_{23} - \rho_{31}}{m_2} & \ \frac{\rho_{31} + \rho_{23} - \rho_{12}}{m_3} & 2\,\frac{\rho_{23}}{m_{23}}
 \end{array}
               \right| \ ,
\end{equation}
is linear in $\rho$-coordinates, cf.(\ref{gmn33-red}). Its determinant is
\[
  D_m\ =\ \text{Det}\big[g^{\mu \nu}({\rho})\big] \ =\ 2\,\frac{m_1+m_2+m_3}{m_1^2\,m_2^2\,m_3^2} \times
\]
\begin{equation}
\label{gmn33-rho-det-M}
 \left(m_1m_2\rho_{12}+m_1m_3\rho_{13}+m_2m_3\rho_{23}\right)
                     \left(2\rho_{12}\rho_{13} + 2 \rho_{12}\rho_{23} + 2 \rho_{13}\rho_{23}-\rho_{12}^2- \rho_{13}^2 - \rho_{23}^2\right) \ ,
\end{equation}
and it is positive definite. It is worth noting a remarkable factorization property of the determinant
\[
D_m \ = \ 32\, \frac{{(m_1+m_2+m_3)}^2}{m_1^2\,m_2^2\,m_3^2}\, I \ S^2_{\triangle} \ = \ \frac{64}{3}\, \frac{{(m_1+m_2+m_3)}^2}{m_1^2\,m_2^2\,m_3^2}\, P_m \ S_m \ .
\]
Hence, $D_m$ is a polynomial function of the $\rho$-variables and it depends on two coordinates alone, namely the product of $I$ and ${S}_{\triangle}^2$.

The three dimensional $\rho$-space with metric (\ref{gmn33-rho}) is not flat.

\subsection{Reduced Hamiltonian at zero angular momentum: $\rho$-representation}

For the special case of vanishing angular momentum $p_{{}_\Om}=0$, the Hamiltonian (\ref{Hredrho}) reduces to
\begin{equation}
\label{Hredrhozero}
\begin{aligned}
  {\cal H}_0 \  & = \ 2\,\bigg[\,\frac{\rho_{12}\,P_{12}^2}{m_{12}} \ + \ \frac{\rho_{23}\,\,P_{23}^2}{m_{23}} \ + \ \frac{\rho_{31}\,\,P_{31}^2}{m_{31}} \ + \ \frac{\rho_{12} + \rho_{31}- \rho_{23}}{m_1}\,P_{12}\,P_{31}\ + \
\\ &
\qquad \quad \frac{\rho_{12} + \rho_{23}- \rho_{31}}{m_2}\,P_{12}\,P_{23} \ + \  \frac{\rho_{23} + \rho_{31}- \rho_{12}}{m_3}\,P_{23}\,P_{31}\,\bigg] \ + \ V(\rho_{12},\,\rho_{23},\,\rho_{31}\,) \ .
\end{aligned}
\end{equation}
In this case, the coefficients in the kinetic energy are linear in the $\rho$-coordinates.
This classical Hamiltonian is in complete agreement with the result obtained through the procedure of \emph{dequantization} from the quantum one, see Ref.\cite{TME3-d}.
In the next section the above result will be generalized to the non planar case $d>2$. Hereafter, the form (\ref{Hredrhozero}) of the reduced Hamiltonian will be used throughout the text. It is the central object to explore in the present consideration.

\section{Three-body system: non-planar case $d>2$}
\label{non planar case}

\subsection{$\rho$-representation}

The extension of the symmetry reduction originally proposed by Murnaghan for planar case $d=2$
to the non-planar case $d=3$ (from a 18-dimensional phase space to 8-dimensional one) is presented in Eq.$(55_2)$ in Ref.\cite{Kampen}. In this case, the reduced Hamiltonian depends on eight dynamical coordinates, namely, the three mutual distances ($r_{12},\,r_{23}$,\,$r_{31}\,$), an angular variable $\om$ and their associated canonical conjugate momenta. However, at zero angular momentum the $\om-$dependent terms disappear and the system is described by the same reduced Hamiltonian (\ref{Hredrhozero}). In the $\rho$-representation it can be proved that this result is valid in any dimension $d > 1$.

\begin{Theorem}
The $3-$body system in $d$-dimensions ($d \geq 2$) with potential of the form $V=V(\rho_{12},\,\rho_{23},\,\rho_{31}\,)$ and zero angular momentum is described by a six-dimensional reduced Hamiltonian ${\cal H}_0$ (in $\rho$-representation)
\begin{equation}
\label{H0d}
  {\cal H}_0 \ = \ {\cal T}_0 \  + \ V(\rho_{12},\,\rho_{23},\,\rho_{31}\,) \ ,
\end{equation}
where the kinetic energy ${\cal T}_0$ is polynomial in coordinates $\rho_{ij}\equiv r_{ij}^2$ and their conjugate momenta $P_{ij}$. Explicitly,
\begin{equation}
\label{}
\begin{aligned}
{\cal T}_0 \ = & \
2\,\bigg[\,\frac{\rho_{12}\,P_{12}^2}{m_{12}} \ + \ \frac{\rho_{23}\,\,P_{23}^2}{m_{23}} \ + \ \frac{\rho_{31}\,\,P_{31}^2}{m_{31}} \ + \ \frac{\rho_{12} + \rho_{31}- \rho_{23}}{m_1}\,P_{12}\,P_{31}\ + \
\\ &
\qquad \quad \frac{\rho_{12} + \rho_{23}- \rho_{31}}{m_2}\,P_{12}\,P_{23} \ + \  \frac{\rho_{23} + \rho_{31}- \rho_{12}}{m_3}\,P_{23}\,P_{31}\,\bigg] \ .
\end{aligned}
\end{equation}

\end{Theorem}

\medskip\noindent {\bf Proof}: Take free 3-body Hamiltonian (kinetic energy)
\begin{equation}
\label{TH}
{\cal T} \ = \ \frac{1}{2\,m_1} \,{\bf p}^2_1 \ + \ \frac{1}{2\,m_2} \,{\bf p}^2_2 \ + \ \frac{1}{2\,m_3} \,{\bf p}^2_3 \ .
\end{equation}
(${\bf p}_i\,=\,m_i\,\dot {\bf r}_i$). Let us make a canonical transformation (change of variables in the phase space) from $({\bf r}_i,{\bf p}_i)$ variables to the $d-$dimensional center-of-mass variable, the three coordinates (squares of mutual distances) $\rho_{ij}$, a suitable set of $(2d-3)$ angular variables and their corresponding canonical momenta. The center-of mass-variables can be separated out completely. The $\rho$-variables are given by
\[
     \rho_{\ell k}\ = \ \sum_{s=1}^d(x_{\ell,s}-x_{k,s})^2,\quad 1\le \ell<k\le 3\ .
\]
Thus,
\[
p_{x_{\ell,s}}\ = \ 2\,\sum_{h\ne \ell,}(x_{\ell,s}-x_{h,s})\,P_{ \rho_{\ell h}} \  + \ \cdots \ ,
\]
where the non-explicit terms are proportional to the momentum variables associated with the remaining angular coordinates. At zero total angular momentum, such terms vanish. Hence, we can compute the coefficient of $P_{ \rho_{\ell k}}^2$
in (\ref{TH}) for $\ell<k$. It is
\[
2 \,\bigg(\frac{1}{m_\ell}\ + \ \frac{1}{m_k}\bigg)\sum_{s=1}^d(\,x_{\ell,s}-x_{k,s}\,)^2 \ = \
2\,\frac{m_\ell+m_k}{m_\ell m_k}\rho_{\ell k}\ .
\]
Similarly the coefficient of  $P_{ \rho_{\ell k}}\,P_{ \rho_{\ell {k'}}}$ for $k< k'$ is
\[
\frac{ 4}{m_\ell}\sum_{s=1}^d(x_{\ell,s}-x_{k,s})(x_{\ell,s}-x_{{k'},s})
=\frac{4 }{m_\ell}({\bf r}_\ell-{\bf r}_k)\cdot({\bf r}_\ell-{\bf r}_{k'})=\frac{2}{m_\ell}(\rho_{\ell {k}}+\rho_{\ell {k'}}-\rho_{k {k'}})\ ,
\]
where the last equality follows from the law of cosines. The coefficient of
$P_{ \rho_{\ell k}}P_{ \rho_{{\ell'} {k'}}}$ for $k,{k'},\ell,{\ell'}$ all
pairwise distinct is $0$. At zero center-of-mass momentum (at rest frame)
we arrive at ${\cal T}_0$.
\quad $\Box$

Hence, for any dimension $d >1$ all trajectories with zero angular momentum are governed by the Hamiltonian (\ref{Hredrhozero}). This Hamiltonian also describes a three-dimensional particle moving in a curved space subject to a potential $V$.

\subsection{Three-body closed chain of interactive harmonic oscillators}

As an application of the presented formalism, we consider the case
of 3-body oscillator with quadratic potentials of interaction which depend on relative {\it distances}, $|{\bf r}_i - {\bf r}_j |$, only \footnote{For quantum counterpart of the is problem, see \cite{3bodychain}.}. Needless to say that the two-body harmonic oscillator can be reduced to a one-dimensional radial Jacobi oscillator, see e.g. \cite{3bodychain}. In the 3-body case such a reduction is not possible in general.

The potential of a three-body closed chain of harmonic oscillators takes the form
\begin{equation}
\label{V3-es}
   V^{(\text{har})}\ =\ 2\,\om^2\bigg[   \nu_{12}\,\rho_{12} \ + \  \nu_{13}\,\rho_{13} \ + \  \nu_{23}\,\rho_{23} \bigg]\ ,
\end{equation}
where $\om > 0$ is frequency and $\nu_{12}, \nu_{13}, \nu_{23} \geq 0$ define spring constants.
% {\color{blue}where $\nu_{12}$, $\nu_{13}$ and $\nu_{23}$ are strictly positive real parameters that % govern the corresponding spring constants.}
At zero total angular momentum in center-of-mass inertia frame, this system is naturally described by the Hamiltonian
\begin{equation}
\label{Hhar}
\begin{aligned}
{\cal H}_0^{\text{har}} \ = & \
2\,\bigg[\,\frac{\rho_{12}\,P_{12}^2}{m_{12}} \ + \ \frac{\rho_{23}\,\,P_{23}^2}{m_{23}} \ + \ \frac{\rho_{31}\,\,P_{31}^2}{m_{31}} \ + \ \frac{\rho_{12} + \rho_{31}- \rho_{23}}{m_1}\,P_{12}\,P_{31}\ + \
 \frac{\rho_{12} + \rho_{23}- \rho_{31}}{m_2}\,P_{12}\,P_{23}
\\ &
\qquad \quad \ + \   \frac{\rho_{23} + \rho_{31}- \rho_{12}}{m_3}\,P_{23}\,P_{31}\,\bigg]\  + \  2\,\om^2\bigg[   \nu_{12}\,\rho_{12} \ + \  \nu_{13}\,\rho_{13} \ + \  \nu_{23}\,\rho_{23} \bigg] \ ,
\end{aligned}
\end{equation}
cf.(\ref{Hredrhozero}). Unlike the case of the standard three-dimensional harmonic oscillator the Hamiltonian (\ref{Hhar}) is not superintegrable and it does not admit separation of variables. However, in the special case
\begin{equation}
\label{super-int}
   m_2\,\nu_{13}=m_3\,\nu_{12}\ ,\qquad m_1\,\nu_{23}=m_2\,\nu_{13}\ ,\qquad m_3\,\nu_{12}=m_1\,\nu_{23}\ ,
\end{equation}
(any two relations imply that the third relation should hold), the system (\ref{Hhar}) admits 5 functionally independent constants of the motion, so it becomes \emph{maximally} superintegrable \cite{3bodychain}. If one of the three constraints in (\ref{super-int}) is fulfilled only the system becomes \emph{minimally} superintegrable.

\section{$(P,S,T)$-representation {(geometrical representation)}}
\label{PST}

In this Section, we consider the case of three equal masses $m_1=m_2=m_3=1$. By using a canonical transformation, the geometrical properties of the triangle of interaction are translated to a certain dynamical variables (see below).

\subsection{Reduced Hamiltonian}

Based on the $\mathbb{Z}_2^{\otimes 3} \oplus \mathcal{S}_3$ symmetry of the free Hamiltonian ${\cal H}_0$ (\ref{Hredrhozero}), let us introduce its invariants as new variables
\[
 (r_{12},\,r_{23},\,r_{31}) \quad \rightarrow \quad
(\rho_{12},\,\rho_{23},\,\rho_{31}) \quad \rightarrow \quad  (P,\,S,\,T) \ ,
\]
where
\begin{equation}
\label{GVa}
\begin{aligned}
&  P \ \equiv  \ \frac{3}{2}\,I \ = \  \frac{1}{2}( \,\rho_{12} \ + \ \rho_{23} \ + \ \rho_{31} \,   ) \ ,
\\ &
 S \ \equiv \ S^2_{\bigtriangleup} \ = \  \frac{1}{16}\big(\,  2\,\rho_{12}\,\rho_{23} \ +\  2\,\rho_{12}\,\rho_{31} \ +\  2\,\rho_{23}\,\rho_{31} \ -\  \rho_{12}^2 \ -\  \rho_{23}^2 \ -\  \rho_{31}^2 \,\big) \ ,
\\ &
T \ \equiv \  \rho_{12}\,\rho_{23}\,\rho_{31} \ ,
\end{aligned}
\end{equation}
cf.(\ref{sigv}), and the three associated canonical conjugate momentum $P_P$, $P_S$ and $P_T$, respectively.

In (\ref{GVa}), the moment of inertia $I \propto P$ and the area (squared) $S^2_{\bigtriangleup}=S$ of the triangle of interaction are translated to two new dynamical variables explicitly. The quantities $P$ and $S$ were called \emph{volume variables}. It is worth mentioning that each of the variables $(P,S,T)$ (\ref{GVa}) is characterized by an \emph{accidental} permutation symmetry $\mathcal{S}_3$ in $\rho-$coordinates in addition to the symmetry $\mathcal{S}_3$ of interchange of any pair of bodies positions.

In new variables the Hamiltonian ${\cal H}_0$ (\ref{Hredrhozero}) takes the form
\begin{equation}
\label{Hredgeo}
\begin{aligned}
  {\cal H}_{\rm geo} \  & = \ 3\,P\,P_P^2 \ + \ P\,S\,P_S^2 \ + \
  {T} \,\bigg[ 48\,S \ + \ 4\,P^2\,  \bigg]\,P_{T} ^2
\\ &
 \ + \ 18\,{T} \,P_P\,P_{T}  \ + \ 12\,S\,P_P\,P_S \ + \ \bigg[\,32\,S^2 \ + \ 8\,S\,P^2 \bigg]\,P_{T} \,P_S
 \ + \ V(P,S,T)  \ .
\end{aligned}
\end{equation}
It corresponds to 3D solid body motion in an external potential $V$. This Hamiltonian ${\cal H}_{\rm geo}$ can be interpreted as the way to describe a three-dimensional particle in a curved space, see below.
Tensor of inertia can be identified with cometric in this case.
The associated cometric for (\ref{Hredgeo}) is given by
\begin{equation}
\label{gmn33-geo}
 g_{}^{\mu \nu}(P,\,S,\,{T})\ = \left|
\begin{array}{ccc}
 3\,P\ & \ 6\,S \ & \ 9\,{T} \\
            &                                   &          \\
  6\,S\ & \  S\,P \ & \ 4 S\,(4\,S \ + \ P^2) \\
            &  \                                &          \\
 9\,{T}\ & \ 4S(4S \ + \ P^2)\ &\ 4\,(12\,S \ + \ P^2)\,{T}\,
\end{array}
               \right| \ .
\end{equation}
Its components are polynomials in the volume variables. Its determinant admits factorization to three factors,
\[
  D_{\rm geo}\ \equiv\ \text{Det}\big[\,g_{}^{\mu \nu}(P,S,T)\,\big]\ =\ 3\,P\,S\,\left(4\, P\, {T}\, \left(36 \,S \ +\ P^2 \right)\ - \ 16\, S\, \left(4\, S\ + \ P^2 \right)^2 \ - \ 27\, {T}^2 \right) \ .
\]

Now, let us study under what conditions the determinant vanishes, $D_{\rm geo}=0$, so that the metric (tensor of inertia) becomes degenerate. In this case, the transformation (\ref{GVa}) is singular (not-invertible). There are three possibilities. The first possibility is when the first factor vanishes, $P=0$. We discard it, since it corresponds to the situation when the configuration space shrinks to a point (we call it {\it triple collision point}).
The second possibility is
\begin{equation}
\label{G203a}
S=0\ ,
\end{equation}
it corresponds to a {\it collinear three-body configuration}, or, equivalently, when three bodies are on the line, the configuration space becomes two-dimensional, thus, it shrinks to plane, while the third possibility leads to,
\begin{equation}
\label{G203b}
   4\,P\,{T}\, \left(36 \,S \ + \ P^2 \right)\ - \ 16\, S\, \left(4\, S\ + \ P^2 \right)^2 \ - \ 27\, {T}^2  \ = \ 0 \ ,
\end{equation}
in which l.h.s. is homogeneous polynomial in $\rho-$coordinates. From (\ref{G203b}) it follows that the determinant $D_{\rm geo}$ vanishes for any isosceles triangle of interaction, in particular, for equilateral triangle, where
\begin{equation}
\label{DES}
      12\,S \ = \ P^2  \qquad , \qquad  27\,{T} \ = \ 8\,P^3   \ .
\end{equation}

The space with metric (\ref{gmn33-geo}) is not flat.

\vspace{0.2cm}

\subsection{Equations of motion}

\vspace{0.2cm}

From the reduced Hamiltonian (\ref{Hredgeo}), we obtain the Newton equations of motion for the variables $(P,S,T)$

\begin{equation}
\label{EQMT}
\begin{aligned}
 \ddot{P} \ = \ & \frac{3}{2\,D_{\rm geo}}\bigg[\, -4 \,S \,\dot{P}^2\, \left(4\, S \,\left(4 S+P^2\right)^2-P \,T \left(12\, S+P^2\right)\right) \ + \
\\ &
3 \,\dot{S}^2\, T\, \left(48\, S\, P+4 \,P^3-27\, T\right) \ - \ 3\, S\, \dot{T}^2 \,\left(12 \,S-P^2\right) \ - \ 12\, S\, \dot{S} \,\dot{P}\,T\, \left(24\, S\, -2\, P^2\right) \ + \
\\ &
\dot{T} \left(6 \,S\,\dot{P} \,\left(8\, S \left(4 \,S+P^2\right)-3\,T\right)-12\, S\, \dot{S}\, \left(8\, S \,P+2\, P^3-9\, T\,\right)\right)\,\bigg] \ - \
\\ &
6\,(3 \,T\,\pa_{T} V  \,+\,2\, S\, \pa_{S} V \,+\,P\, \pa_{P} V ) \ ,
\end{aligned}
\end{equation}
\begin{equation}
\label{EQMS}
\begin{aligned}
 \ddot{S} \ = \ & \frac{3}{2\,D_{\rm geo}\,P}\bigg[  \dot{S}^2 \,\left(4\, P\, T \left(-72\, S^2+30\, S\, P^2+P^4\right)-16\, S \,P^2 \left(4 \,S+P^2\right)^2+27\, T^2 \,\left(6\,S-P^2\right)\right)  \ + \
\\ &
 6 \,S^2\, \dot{T}^2\, \left(12\, S-\, P^2\right) \ + \ 8 \,S^2\, \dot{P}^2 \,\left(4 \,S\, \left(4 \,S+P^2\right)^2-P\,T \left(12\, S+P^2\right)\right) \ + \
\\ &
2\, S\, \dot{S}\, \dot{P} \,\left(4\, T \,\left(72 \,S^2+30 \,S \,P^2+P^4\right)-16\, S\, P\, \left(4 \,S+P^2\right)^2-27\, P\, T^2\right) \ + \
\\ &
\dot{T}\, \left(24 \,S^2\, \dot{S}\, \left(8\, S\, P+2\, P^3-9\, T\right)-12\, S^2\, \dot{P}\, \left(8 \,S \left(4\, S+P^2\right)-3\, P\, T\,\right)\right)  \,\bigg]
\\ &
 \ - \ 2\,S\,( 4\, \left(4 \,S+P^2\right) \,\partial_{T} V\,+\,P\, \partial_{S} V\,+\,6\, \partial_{P} V  ) \ ,
\end{aligned}
\end{equation}
\begin{equation}
\label{EQMV}
\begin{aligned}
\ddot{T} \ = \ & \frac{3}{2\,D_{\rm geo}\,P}\bigg[ 8 \,S \,\dot{P}^2\, T \left(96\, S^3-48\, S^2\, P^2+14\, S\, P^4-3 \,P^3\, T\right) \ + \
 \\ &
\dot{S}^2 \,T \,\left(16 \,P^2\, \left(96\, S^2+12 \,S\, P^2+P^4\right)-144\, P\, T\, \left(6 \,S+P^2\right)+243 \, T^2\right) \ + \
\\ &
S \,\dot{T}^2 \,\left(-192 \,S^2\, P+12\, S\, \left(8\, P^3+9\, T\right)+4\, P^5-45 \,P^2 \,T\right) \ - \
\\ &
8 \,S\, \dot{S}\, \dot{P}\, T\, \left(192\, S^2\, P-12\, S\, \left(8\, P^3+9\, T\right)-4 \,P^5+45\, P^2\, T\right)
\\ &
\ + \
\dot{T} \bigg(\,16\, S \,\dot{P} \,\left(32\, S^3\, P-4\, S^2\, \left(4 \,P^3+9\, T\right)-3\,S\,P^2\, \left(2\, P^3-5\, T\right)+P^4\, T\right) \ - \
\\ & 4 \,S \,\dot{S}\, \left(-6 \,P\, T\, \left(36\, S+13\, P^2\right)+8\, P^2\, \left(4 \,S+P^2\right) \left(12\, S+P^2\right)+81\, T^2\right)\,\bigg)  \,\bigg]
\\ &
 \ - \ 2\,(\,4\, T\, \left(12\, S+P^2\right) \partial_{T} V\,+\,4\, S\, \left(4\, S+P^2\right) \partial_{S} V\,+\,9\, T \,\partial_{P} V\,) \ .
\end{aligned}
\end{equation}
A further remark is in order. From the Hamiltonian (\ref{Hredgeo}) it follows that the time evolution of $P_T$ is given by the Hamilton's equation
\begin{equation}
\label{PTdinEv}
  \dot{P}_T \ = \  -2\,P_T\,\big[\, 9\,P_P \ + \ 2\,P_T\,(12\,S+P^2) \big]\ -\ \pa_T V\ .
\end{equation}
Note when potential $V$ depends on the volume variables $P$ and $S$ only, (\ref{PTdinEv}) is reduced to
\begin{equation}
\label{}
  \dot{P}_T \ = \  -2\,P_T\,\big[\, 9\,P_P \ + \ 2\,P_T\,(12\,S+P^2) \big]\ ,
\end{equation}
which implies the existence of trajectories with $P_T=0$. From the Hamilton's equations we obtain
\begin{equation}
\label{PTvel}
  P_T \ = \ \frac{3\,S}{2\,D_{\rm geo}}\bigg[ {\dot{P}} \left(8 S \left(P^2+4 S\right)-
  3 P T\right)+{\dot{S}}  \left(18 T-4 P \left(P^2+4 S\right)\right)+{\dot{T}}
  \left(P^2-12 S\right)\,\bigg] \ .
\end{equation}
Therefore, one can construct a \emph{reduced} Hamiltonian on the phase space $(P,S,P_P,P_S)$. This construction will be elaborated in the next Section.

Now, within the ($P,S,T$)-representation we consider two particular three-body examples: (i) the $\mathbb{R}^3$ Newtonian gravity and (ii) the planar, $\mathbb{R}^2$ choreographic trajectories on the algebraic lemniscate by Jacob Bernoulli (1694).

\subsubsection{Three-body Newtonian gravity potential}

First, let us consider the three-body Newton problem in $\mathbb{R}^3$ ($d=3$). The potential in (\ref{Hredgeo}) reads
\begin{equation}
\label{3-body-Newton}
   V \ \equiv \ V_\gamma \ =\ - \gamma \,\bigg( \frac{1}{r_{12}} \ + \ \frac{1}{r_{23}}\ + \   \frac{1}{r_{31}} \bigg) \ =\ - \gamma \,\bigg( \frac{1}{\sqrt{\rho_{12}}} \ + \  \frac{1}{\sqrt{\rho_{23}}}\ + \ \frac{1}{\sqrt{\rho_{31}}} \bigg) \ ,
\end{equation}
where $\gamma$ is the gravitational constant. In terms of the variables ($P,S,T$), see (\ref{GVa}),
one can show that $ V_\gamma$ is one of the roots of the fourth order algebraic equation
\begin{equation}
\label{V4g}
  T^2\,V_\gamma^4 \ - \ 2\,\gamma^2\,(4\,S+P^2)\,T\,V_\gamma^2 \ {+} \ 8\,\gamma^3\,{T}^{\frac{3}{2}}\,V_\gamma \ + \ \gamma^4\,\bigg({(4\,S+P^2)}^2 \ -\ 8\,{T}\,P\bigg) \ = \ 0 \ .
\end{equation}
This equation is invariant with respect to simultaneous change $(\gamma \rar -\gamma)$ and $(V_\gamma \rar -V_\gamma)$ \footnote{For the case of $\mathbb{R}^2$ ($d=2$) Newtonian gravity the 3-body potential is of logarithmic type, $V_{\gamma}~=~\gamma \log r_{12} r_{13} r_{23}$.
Analogue of the equation (\ref{V4g}) is very simple, $V_{\gamma}=\frac{\gamma}{2}\, \log T$.
It does not contain dependencies on $P,S$}.

Interestingly, the remaining three roots correspond to three different 3-body Coulomb potentials for 3 unit charges of different signs, $\pm i$,

\begin{equation}
\label{V4g-3roots}
\begin{aligned}
& V^{(1)}_\gamma  \ = \ \gamma \,\bigg(\, -\frac{1}{\sqrt{\rho_{12}}} \ + \  \frac{1}{\sqrt{\rho_{23}}}\ + \ \frac{1}{\sqrt{\rho_{31}}}\, \bigg)\ ,
\\ &
V^{(2)}_\gamma  \ = \ \gamma \,\bigg(\, \frac{1}{\sqrt{\rho_{12}}} \ - \  \frac{1}{\sqrt{\rho_{23}}}\ + \ \frac{1}{\sqrt{\rho_{31}}} \,\bigg)\ , 
\\ &
V^{(3)}_\gamma  \ = \  \gamma \,\bigg(\, \frac{1}{\sqrt{\rho_{12}}} \ + \  \frac{1}{\sqrt{\rho_{23}}}\ - \ \frac{1}{\sqrt{\rho_{31}}} \,\bigg)  \ .
\end{aligned}
\end{equation}

\vspace{0.2cm}

Hence, the above equation (\ref{V4g}) describes all four possible 3-body Newton/Coulomb potentials with imaginary unit charges with constant of interaction $\gamma>0$.

The discriminant of the equation (\ref{V4g}) admits factorization
\begin{equation}
\label{DetVC}
 D_{\gamma}\ =\ 4096 \,\gamma^{12}\, {T}^8\, \left(4\, P\, {T}\, \left(36 \,S \ + \ P^2\right)\ - \ 16\, S\, \left(4\, S\ + \ P^2\right)^2 \ - \ 27\, {T}^2\right)\ ,
\end{equation}
and its last factor coincides with the last factor in the factorized expression for determinant $D_{\rm geo}$ of the metric (\ref{gmn33-geo}). For any isosceles triangle both determinant and discriminant vanish, $D_{\gamma}=D_{\rm geo}=0$.
Furthermore, in the case of an equilateral triangle of interaction the potential (\ref{3-body-Newton}) simplifies,
\[
 V_{\gamma} \ = \ -\frac{3^{\frac{3}{2}}\gamma}{\sqrt{2P}}\ .
\]
It corresponds to the solution found long ago by Lagrange \cite{Lagrange}, it describes to the so called {\it central configuration}, see e.g. Refs.\cite{Moeckel, Saari, Albouy} and reference therein. Also, the remarkable periodic Figure Eight solution of the three-body problem found by Moore \cite{Moore} and confirmed by Chenciner and Montgomery \cite{Chenciner-Montgomery} - the so-called 3-body choreography - gets a natural presentation in the geometrical variables. On this trajectory, the motion of the bodies alternates between six collinear configurations and six isosceles-triangular ones.

\subsubsection{Choreographic motion}

As for the choreographic trajectories in $\mathbb{R}^2$ ($d=2$), when three bodies follow one to each other being on the same curve, Fujiwara et al. \cite{Fujiwara} solved the inverse problem of choreographic motion on a Figure-8 given by the algebraic lemniscate of Jacob Bernoulli (1694) on $(x,y)$-plane
\begin{equation}
\label{lemn}
 (x^2\,+\,y^2)^2\ = \ c^2 \,(x^2-y^2)\ ,
\end{equation}
where without loss of generality one can put $c=1$, and found 3-body pairwise potential. It was shown that such a figure eight is the choreographic trajectory for three unit mass, point-like particles with zero angular momentum with two-parametric potential
\begin{equation}
\label{potential}
 V_\infty\ =\ \frac{1}{4}\, \ln {T} - \frac{\sqrt{3}}{12}\, P\ ,
\end{equation}
with the {volume} variables $P,T$ given by (\ref{GVa}), \cite{ATJC}.
The first attractive terms in (\ref{potential}) represents a 3-body $\mathbb{R}^2$ Newtonian potential for gravitational constant $\gamma=1/2$, while the second repulsive term is the square of the hyperradius in the space of relative motion. It is worth mentioning that the Figure-8 algebraic lemniscate with potential (\ref{potential}) is \emph{close} to the transcendental Figure-8 trajectory found by C Moore\cite{Moore}, see also Ref.\cite{Chenciner-Montgomery}, for the $\mathbb{R}^3$ Newtonian gravity potential.

\begin{figure}[htp]
  \centering
  \includegraphics[width=8.0cm]{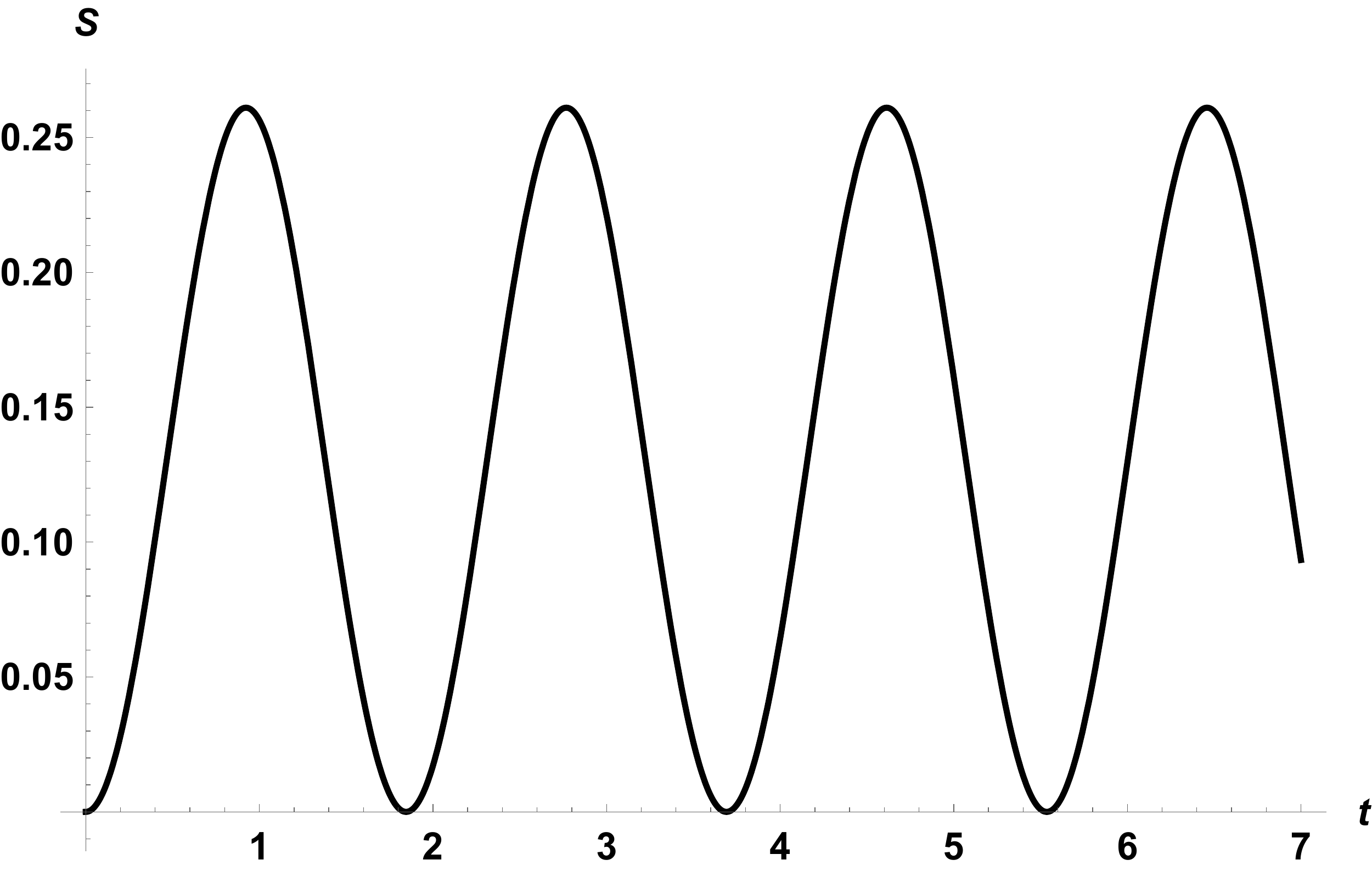}
  \caption{Time evolution of the volume variable $S$, area (squared) of the triangle of interaction, for the 3-body choreographic Figure-8 on the algebraic lemniscate by Bernoulli. The points $S=0$ correspond to the so-called Euler line for which all three bodies are on the line.}
  \label{elliptic}
\end{figure}

Recently, in Ref.\cite{ATJC} it was found that the 3-body choreographic motion on the algebraic lemniscate is maximally (particularly) superintegrable. Moreover, the two variables $P$ and $T$ become particular integrals (see also Ref.\cite{Vieyra}). Along this Figure-8 trajectory, these variables take the constant value
\begin{equation}
\label{CTV}
  P \ = \ {T} \ = \ \frac{3\,\sqrt{3}}{2} \ .
\end{equation}
Therefore, it is natural to work with the Hamiltonian $(\ref{Hredgeo})$. It can be shown that three equations of motion (\ref{EQMT})-(\ref{EQMV}) constrained by the conditions (\ref{CTV}) lead to non-linear ODE
\begin{equation}
\label{Weierstrasse}
 S \,\left(\,256\, S^2\,+\,864\, S\,-\,243\,\right) \ + \ 48\, \sqrt{3}\, \dot{S}^2 \ = \ 0 \ ,
\end{equation}
which defines the elliptic curve, see Fig.\ref{elliptic}, with variables $P,T$ playing role of elliptic invariants \cite{ATJC}. The solution of (\ref{Weierstrasse}) is given by the Weierstrass function
$\wp (t; P,T)$.
Hence, in the configuration space parametrized by geometrical variables the 3-body choreography on algebraic lemniscate by Jacob Bernoulli is given by a (planar) elliptic curve!

\section{Volume variables representation}
\label{volumeR}

\subsection{Case of equal masses}

\label{EMcase}

{

Now we consider the case of three unit masses $m_1=m_2=m_3=1$ and focus on potentials that depend solely on the volume variables $P$ and $S$, see (\ref{GVa}),
\begin{equation}
\label{VVolV}
      V \ = \ V(P,\,S) \ .
\end{equation}
In this case the equations of motion (\ref{PTdinEv}) that emerge from the Hamiltonian ${\cal H}_{\rm geo}$ (\ref{Hredgeo}) admit a number of solutions with $P_T=0$. This fact is not trivial since the momentum $P_T$ is not a Liouville integral of motion, $\{P_T,\,{\cal H}_{\rm geo}\}_{P.B.}\neq 0$.

Equivalently, in the six-dimensional phase space $(P,S,T,P_P,P_S,P_T)$ the hypersurface $P_T=c$
becomes an invariant manifold \cite{Wiggins} at $c=0$: $\{P_T,\,{\cal H}_{\rm geo}\}_{P.B.}|_{{}_{P_T=0}} \ =\ 0$. This implies that any trajectory of ${\cal H}_{\rm geo}$ for which the initial condition $P_T=0$ is imposed will remain on the hypersurface ${\cal H}_{\rm geo}(P,S,T,P_P,P_S,P_T=0)=E$ during the evolution.
Inserting $P_T=0$ directly into the Hamiltonian (\ref{Hredgeo}) we arrive at
\begin{equation}
\label{HvolVar}
   {\cal H}_{\rm vol}\ \equiv \  {\cal H}_{\rm geo}\mid_{{}_{\small P_T=0}}\ =
    \ 3\,P\,P_P^2 \ + \ P\,S\,P_S^2 \ + \ 12\,S\,P_P\,P_S  \ + \ V(P,S)  \ ,
\end{equation}
which describes a two-dimensional particle in a curve space, see below. Notice that $T-$dependence in (\ref{HvolVar}) is absent as well.

From ${\cal H}_{\rm vol}$ we obtain the Newton equations
\begin{equation}
\label{HamEqN}
\begin{aligned}
& \ddot{P} \ = \ \frac{1}{2\,D_{\rm vol}}\bigg[\, 3\, P \,S\,\dot{P}^2\ -\ 36\, S\,\dot{P} \, \dot{S}\ + \ 9\, P\, \dot{S}^2 \,\bigg] \ - \
6 \left(\,2\, S\, \partial_S \,V \ + \ P \,\partial_P\,V\,\right)
\\ &
\ddot{S} \ = \  \frac{1}{2\,D_{\rm vol}}\bigg[\, 3\, \dot{S}^2 \,\left(P^2-18 S\right)\ - \ 6\,S^2\, \dot{P}^2  \ + \ 6\, P\,S\, \dot{P} \, \dot{S} \,\bigg] \ - \  2 \,S \,\left(\,P \,\partial_S \,V \ + \ 6\, \partial_P \,V \,\right) \ ,
\end{aligned}
\end{equation}
cf.(\ref{EQMT}),(\ref{EQMS}).

\noindent
\textbf{Remark.} It is worth clarifying the connection between the trajectories of ${\cal H}_{\rm vol}$ (\ref{HvolVar}) and those of ${\cal H}_{\rm geo}$ (\ref{Hredgeo}). The evolution $P=P(t)$ and $S=S(t)$ in (\ref{HamEqN}) also satisfies the equations of motion (\ref{EQMT})-(\ref{EQMV}), where $T=T(t)$ is fully determined by $P(t)$ and $S(t)$ if the condition $P_T=0$ is imposed, see (\ref{PTvel}). In this way we reduced effectively the dimensionality of the system (\ref{Hredgeo}) from six to four.

The Hamiltonian (\ref{HvolVar}) is written in what we call the \emph{volume variables representation}.

\subsubsection{The potential $V=\frac{{F[P/ \sqrt{S}]}}{\sqrt{S}}$}

Let us consider an arbitrary potential $V=V(P,S)$ on the submanifold of phase space for which ${\cal H}_{\rm vol}=0$ (zero energy level) and $\dot{P}=0$ (a constant moment of inertia). In this case, from (\ref{HvolVar}) and (\ref{HamEqN}) it follows that the potential must obey the equation
\begin{equation}\label{PMG}
2\,S\,\partial_S\,V \ + \ P\,\partial_P\,V \ = \ -V \ .
\end{equation}
The solution of (\ref{PMG}) is given by
\[
V\ =\ \frac{{U[P/ \sqrt{S}]}}{\sqrt{S}} \ ,
\]
where $U[z]$ is an arbitrary function of the argument $z$. The particular case $U_0=-\gamma \frac{P}{\sqrt{S}}$, thus $V=-\gamma \frac{P}{S}$ has been fully analyzed in Ref.\cite{Montgomery} where the author constructs the hyperbolic plane with its geodesic flow as the scale plus symmetry reduction of a three-body problem in the Euclidean plane. In fact, in Ref.\cite{Montgomery} the generalization of $V=-\gamma \frac{P}{S}$ to the case of arbitrary masses was considered as well. Later on, we will show that the mass-dependent case can be easily obtained from the present equal-mass case.

\vspace{0.2cm}

For an arbitrary function $U[P/ \sqrt{S}]$ we obtain the first-order nonlinear ODE for the volume variable $S$
\begin{equation}\label{}
P\,{\dot{S}}^2 \ + \ 4\,U\,(P^2 \ - \ 12\,S)\,\sqrt{S} \ = \ 0 \ ,
\end{equation}
where $P>0$ plays the role of an external parameter. In the case of the potential $V=-\gamma \frac{P}{S}$ we arrive to the solution
\begin{equation}\label{}
  S(t) \ = \ \frac{P^2}{12} \ - \ 3\, \gamma\,{(c_1\ + \ 2\, t)}^2 > 0 \ ,
\end{equation}
here $c_1$ is a constant of integration. Notice that the potential $V \propto \frac{1}{P}$ leads to the solutions $S(t)=12\,P^2$ (an equilateral triangle) and $S(t)=0$ (three particles on a line).

\subsubsection{The metric $g_{\rm vol}^{\mu \nu}$}

Now, the metric (or, equivalently, the tensor of inertia) for (\ref{HvolVar}) is of the form
\begin{equation}
\label{VGemVol}
 g_{\rm vol}^{\mu \nu}\ = \left|
 \begin{array}{cc}
 3\,{P} & \ 6\,S  \\
  6\,S & P\,S     \,
 \end{array}               \right| \ ,
\end{equation}
with determinant
\[
  D_{\rm vol}\ = \ 3\,S\,(\,{P}^2 \ - \ 12\,S\,)\ .
\]
In the case of an \emph{equilateral triangle of interaction} the relation ${P}^2 \ = \ 12\,S$ holds and this determinant vanishes, ${D}_{\rm vol}=0$.
It also follows from (\ref{VGemVol}) that the corresponding Ricci scalar $\rm {Rs}$ is given by
\begin{equation}
\label{Rs1}
 {\rm  Rs}_{\rm vol} \ = \   \frac{3\,S\,P\,(\, S \ - \ 3 \,)}
     {{D_{\rm vol}}^2} \ ,
\end{equation}
see below, (\ref{Rsm}). {${\rm  Rs}_{\rm vol}$} is singular for equilateral triangle of interaction.

{
An interesting particular case occurs when the potential (\ref{VVolV}) does not depend on $S$ and is a function of the volume variable $P$ alone
\[
V  \ = \ V(P) \ .
\]
It follows from (\ref{HvolVar}) that there exist trajectories for the original Hamiltonian (\ref{Hredgeo}) which depend on $P$ only. Such trajectories lie on the intersection between the hypersurfaces $P_T=0$ and $P_S=0$, they are described by the two-dimensional Hamiltonian
\begin{equation}
\label{HPP}
%\begin{aligned}
 {\cal H}_{\rm vol}\mid_{{}_{\small P_S=0}}\ \equiv \  {\cal H}_P \ =\ 3\,{P}\,P_{P}^2\ +\ V(P)\ ,
%\end{aligned}
\end{equation}
cf.(\ref{HvolVar}). The connection between the trajectories of ${\cal H}_P$ (\ref{HPP}) with those of ${\cal H}_{\rm vol}$ (\ref{HvolVar}) and ${\cal H}_{\rm geo}$ (\ref{Hredgeo}) is the following. Taking a non-trivial solution $P=P(t)\neq 0$ of (\ref{HPP}) one can define the function $S=S(t)$ for which the canonical momentum $P_S$ vanishes in ${\cal H}_{\rm vol}$. Explicitly, the condition can be obtained from (\ref{HvolVar}),
\begin{equation}
\label{PSzero}
  P_S \ = \   \frac{2 \,{\dot P} \,S \,-\,P\,{\dot S}}{2\,S\,( 12\,S\,-\, P^2)} \ = \ 0 \ .
\end{equation}
Such $P=P(t)$ and $S=S(t)=\lambda\,P^2(t)$, with $\lambda \neq 12$ a non-negative parameter, satisfy the equations (\ref{HamEqN}) for ${\cal H}_{\rm vol}$. Moreover, they also obey the equations (\ref{EQMT})-(\ref{EQMV}) for ${\cal H}_{\rm geo}$ where at $\lambda\neq0$ the variable $T$ turns out to be identically zero due to the condition $P_T=0$, hence, the 3body problem degenerates into a 2body system. At $\lambda=0$, thus $S=0$, in general $T\neq0$ and the domain of the problem degenerates to the line.

\subsubsection{Anharmonic Oscillator potential}

As a concrete example, let us consider the following potential
\begin{equation}
\label{VANP}
  V^{(AO)} \ = \ A\,P \ + \ B \,P^2 \ = \ \frac{A}{2}(\, \rho_{12} \, + \,\rho_{23} \, + \,\rho_{31}    \,) \ + \ \frac{B}{4}{(\,\rho_{12}\, + \,\rho_{23}\, + \,\rho_{31}\,)}^2 \ ,
\end{equation}

\noindent
where $B \geq 0$ and $A > 0$ if $B=0$. In terms of variable $P$ the potential (\ref{VANP}) corresponds to a shifted one-dimensional harmonic oscillator on the half line $\mathbb{R}_+$ with non-standard kinetic energy. In the $\rho-$representation it corresponds to a certain three-dimensional quadratic potential on the first octant $\mathbb{R}^3_+$ whilst in the $r-$representation it describes a three-dimensional isotropic harmonic oscillator with quartic anharmonicity \footnote{Note that in present formalism, a more general anharmonic potential
$V(P,S) \ = \ A\,P  + B \,P^2 + C\,S$ (the general $S_3$-symmetric three-variable quartic potential
in $r$-variables) can be studied in straightforward manner.}.

It is worth mentioning that even symmetric harmonic 3body planar systems, with equal spring rest lengths $L$, equal spring constants $k$ and equal masses $m$ despite its apparent simplicity, display a wide array of interesting dynamics for different energy values, (see Ref.\cite{Katz} and references therein). In Ref.\cite{Katz} the authors demonstrate that for such systems the orientation of the triangle is a nontrivial, history-dependent variable of the system, and serves as a sensitive measurable for the type of dynamics the system follows.

For the potential (\ref{VANP}), the Hamiltonian (\ref{HPP}) becomes
\begin{equation}
\label{HPPAO}
\begin{aligned}
  {\cal H}^{(AO)}_P \  & = \ 3\,{P}\,P_{P}^2  \ + \ A\,P \ + \ B \,P^2  \ .
\end{aligned}
\end{equation}

The time-evolution of the system is given by
\begin{equation}
\label{PtSolu}
P(t) \ =\ x^2\ = \ \frac{A\,k^2}{B\,(1-k^2)}\,\text{sn}^2(y,ik) \quad ,\qquad y\,=\, \pm \sqrt{\frac{3\,A}{1-k^2}}\ t\ ,
\end{equation}
($B \neq 0$, $k \neq \pm 1$) where sn$(y,ik)$ is the Jacobi elliptic function with imaginary elliptic modulus $(ik)$. In the case of physical systems the function $P(t)$ should be positive and the parameter $k \neq \pm 1$ is real.

For the solution (\ref{PtSolu}) the Hamiltonian has a meaning of energy and it takes the value
\[
{\cal H}^{(AO)}_P\ = \ \frac{A^2\, k^2}{B \left(1-k^2\right)^2}\ .
\]

Putting the $B = 0$ in (\ref{HPPAO}) we obtain the harmonic oscillator potential $V=A\,P$, which is quadratic polynomial in the $r$-representation. In this case the trajectories are trigonometric vibrations, $P(t)  =   c_1\,\cos^2(\sqrt{3\,A}\,t+c_2)$ with energy ${\cal H}^{(AO)}_P\, =\, \ c_1\,A$ where $c_1>0, c_2$ are real constants of integration. In general, due to the energy conservation ${\cal H}^{(AO)}_P=E$ the trajectories in the phase space are always cubic curves,
\[
   3\,{P}\,P_{P}^2  \ + \ A\,P \ + \ B \,P^2 \ =\ E\ .
\]
Note that at fixed values of the energy $E$ and $A$ the presence of anharmonicity ($B>0$) in (\ref{HPPAO}) tends to decrease the amplitude of the harmonic motion $(B=0)$.

\subsection{Case of unequal masses: modified volume variables}
\label{volumeRR}

In this Section the case of three bodies with arbitrary masses $(m_1,m_2,m_3)$ is considered. From the general Hamiltonian ${\cal H}_0$ at zero angular momentum, see (\ref{Hredrhozero}),
it will be constructed a four-dimensional \emph{reduced} Hamiltonian, which depends on the modified \emph{volume variables} $P_m$ and $S_m$ alone and their respective canonical momenta, for two-variable potentials $V(P_m,\,S_m)$.

As a first step, let us make the change of variables in the Hamiltonian ${\cal H}_0$ (\ref{Hredrhozero})
\begin{equation}\label{CCVV}
(\, \rho_{12},\,\rho_{31},\,\rho_{23} \,) \ \rightarrow  \ (\, P_m,\,S_m,\,Q    \,) \ ,
\end{equation}
where the volume variable
\begin{equation}
\label{Sm}
  S_m \ =\ \frac{3\,m_1\,m_2\,m_3}{m_1+m_2+m_3}\,S \ ,
\end{equation}
is proportional to the square of the area of the triangle of interaction $S$ (\ref{GVa}), and
\begin{equation}
\label{Ptilde}
  P_m \ =\ \frac{1}{2}\bigg[\frac{1}{m_3}\,\rho_{12}\ +\ \frac{1}{m_1}\,\rho_{23}\ + \ \frac{1}{m_2}\,\rho_{31}\, \bigg] \ ,
\end{equation}
is the \emph{weighted sum} of the \emph{edges (squared)} of the triangle of interaction. At equal masses $m_1=m_2=m_3=1$ the modified volume variables coincide to the original ones $P$ and $S$: $P_m \rar P$, $S_m \rar S$. The third variable $Q=Q(\rho_{12},\rho_{31},\rho_{23})$ can be any function of $\rho$'s with condition that the Jacobian of the transformation (\ref{CCVV}) is invertible (nonsingular) in the domain $\mathbb{R}^3_+$, where the problem is defined. In these variables, the Hamiltonian (\ref{Hredrhozero}) takes the form
\begin{equation}
\label{HQ}
\begin{aligned}
    {\cal H}_{Q}\ &\ = \ \frac{m_1+m_2+m_3}{m_1\,m_2\,m_3}\,{P_m}\,P_{{\small P_m}}^2\ + \ P_Q\,(F_1\,P_{{\small P_m}} \ + \ F_2\,P_{S_m} \ + \ F_3\,P_Q)
\\
    &\ +\ \frac{m_1+m_2+m_3}{3\,m_1\,m_2\,m_3}{P_m}\,S_m\,P_{S_m}^2\ +\ 4\,\frac{m_1+m_2+m_3}{m_1\,m_2\,m_3}\,S_m\,P_{{\small P_m}}\,P_{S_{m}}\  + \ V \ ,
\end{aligned}
\end{equation}
where, in general, the coefficients $F_1$, $F_2$ and $F_3$ are functions of $(P_m, S_m, Q)$.
Let us consider that the family of two-variable potentials in (\ref{HQ}) depends on the volume variables
\begin{equation}
\label{V3bg}
      V \ = \ \frac{m_1+m_2+m_3}{3\,m_1\,m_2\,m_3}\,V_m(P_m,\,S_m) \ ,
\end{equation}
alone. In this case it is easy to see that the Hamiltonian (\ref{HQ}) admits trajectories
with $P_Q=0$ in the six-dimensional phase space $(P_m,S_m,Q,P_{P_m},P_{S_m},P_Q)$: $\{P_Q,\,{\cal H}_{\rm Q}\}_{P.B.}\mid_{{}_{P_Q=0}} \ =\ 0$.
Such trajectories are described by the Hamiltonian
\begin{equation}
\label{HredgeoM}
\begin{aligned}
  {\cal H}_m & \  \equiv \ {\cal H}_{0} \mid_{P_Q=0}\\ &
  \  = \ \frac{m_1+m_2+m_3}{3\,m_1\,m_2\,m_3}\bigg[\,3\,{P_m}\,P_{{\small P_m}}^2 \ + \ {P_m}\,S_m\,P_{S_m}^2
 \ + \ 12\,{S_m}\,P_{{\small P_m}}\,P_{S_m}
   \ + \ V_m({P_m},\,{S_m})\,\bigg]  \ ,
\end{aligned}
\end{equation}
which generalizes ${\cal H}_{\rm vol}$ (\ref{HvolVar}) to the non-equal mass case. Multiplying ${\cal H}_m$ (\ref{HredgeoM}) by $\frac{3\,m_1\,m_2\,m_3}{m_1+m_2+m_3}$ we obtain the Hamiltonian (\ref{HvolVar}). Hence, we arrive to the interesting result
\begin{quote}
\emph{The zero total angular momentum trajectories of a 3-body system with equal masses in a two-variable potential $V(P,S)$ and those of 3 bodies with arbitrary masses and potential $\frac{m_1+m_2+m_3}{3\,m_1\,m_2\,m_3}V(P_m,S_m)$ do coincide!}\\
\end{quote}
The Hamiltonian (\ref{HredgeoM}) describes a two-dimensional particle moving in curved space or, equivalently, two-dimensional solid body in external potential $V$. The trajectories of ${\cal H}_m$ are solutions to the equations of motion for ${\cal H}_Q$ (\ref{HQ}) as well.

The corresponding metric for (\ref{HredgeoM}) takes the form
\begin{equation}
\label{VGem}
 g_{m}^{\mu \nu}\ = \left|
 \begin{array}{cc}
 \frac{m_1+m_2+m_3}{m_1\,m_2\,m_3}\,{P_m} & \ 2\,\frac{m_1+m_2+m_3}{m_1\,m_2\,m_3}\,S_m  \\
  2\,\frac{m_1+m_2+m_3}{m_1\,m_2\,m_3}\,S_m \quad & \frac{m_1+m_2+m_3}{3\,m_1\,m_2\,m_3}\,{P_m}\,S_m        \,
 \end{array}
               \right| \ ,
\end{equation}
with determinant
\[
  {\tilde D}_{m}\ = \ \frac{{(m_1 \,+\, m_2 \,+\,m_3)}^2 }{{3\,(m_1\,m_2\,m_3)}^2}\,S_m\,[\,{P_m}^2 \,-\,12\,S_m   \,]
  \ ,
\]
cf.(\ref{VGemVol}). It is worth emphasizing that the metrics (\ref{VGem}) and (\ref{VGemVol}) coincide
up to a multiplicative factor. From (\ref{VGem}) it follows that the corresponding Ricci scalar $\rm {Rs}$ is given by
\begin{equation}
\label{Rsm}
 { \rm \tilde Rs} \ = \   \frac{m_1\,m_2\,m_3\,{P_m}\,(\,S_m\,-\,3)}{(m_1 \,+\, m_2 \,+\,m_3)\,S_m\,{[\,{P_m}^2 \,-\,12\,S_m \, ]}^2} \ = \ \frac{{(m_1 \,+\, m_2 \,+\,m_3)}^3\,S_m\,{P_m}\,(\,S_m\,-\,3)}{9\,{(m_1\,m_2\,m_3)}^3\, {\tilde D}_{m}^2}   \ ,
\end{equation}
cf.(\ref{Rs1}).

In the particular case of a one-variable potential,
\[
   V \ = \ \frac{m_1+m_2+m_3}{3\,m_1\,m_2\,m_3}\,V_m(P_m)  \ ,
\]
as it follows from (\ref{HredgeoM}), the Hamiltonian ${\cal H}_{Q}$ (\ref{HQ}) admits trajectories depending on $P_m$ only. They lie on the intersection between the hypersurfaces $P_Q=0$ and $P_{S_m}=0$, they are governed by the two-dimensional Hamiltonian
\begin{equation}
\label{HredgeoMP}
  {\cal \tilde H}_m  \  \equiv \ {\cal H}_m \mid_{P_{S_m}=0}\ = \ \frac{m_1+m_2+m_3}{3\,m_1\,m_2\,m_3}\,\bigg[\,3\,{P_m}\,P_{{\small P_m}}^2\ +
  \ V({P_m}) \,\bigg] \ ,
\end{equation}
cf.(\ref{HPP}).

As concrete example one can consider the mass-dependent 3-body anharmonic oscillator potential
\[
V_m^{(AHO)}\ = \ A\,P_m\ + \  B\,P_m^2 \ ,
\]
cf.(\ref{VANP}), where $A,B$ are parameters. As mentioned above, it is evident that multiplying the Hamiltonian ${\cal \tilde H}_m$ (\ref{HredgeoMP}) by $\frac{3\,m_1\,m_2\,m_3}{m_1+m_2+m_3}$ we arrive at the same Hamiltonian ${\cal H}_P$ (\ref{HPPAO}). Hence, the trajectories of two 3-body anharmonic oscillators defined by (\ref{HPPAO}) and (\ref{HredgeoMP}), respectively, coincide and, in general, they are given by (\ref{PtSolu}).

In the case of particular 3-body anisotropic harmonic oscillator with different masses which occurs at $B=0$, see above,
\[
  V_m^{(HO)}\ = \ A\, P_m \ =\  \frac{A}{2}\bigg(\,\frac{r_{12}^2}{m_3}\ + \ \frac{r_{31}^2}{m_2}
  \ +\ \frac{r_{23}^2}{m_1}\,\bigg) \ ,
\]
the spring constants in (\ref{V3-es}) take values $\nu_{12}=\frac{1}{4\,\om^2\,m_3}$, $\nu_{13}=\frac{1}{4\,\om^2\,m_2}$ and $\nu_{23}=\frac{1}{4\,\om^2\,m_1}$. It leads to the periodic trigonometric solution
\[
      P_m(t) \ = \  c_1\,\cos^2(\sqrt{3\,A}\,t+c_2) \ ,
\]
where $c_1>0$ and $c_2$ are constants of integration.

\section{Conclusions}
\label{Conclusion}

In this paper, which is the first in a series, the 3-body classical system in a $d$-dimensional coordinate space $\mathbb{R}^d$, $d\geq2$, is considered. The study is restricted to potentials $V=V(r_{12},r_{23},r_{13})$ that depend solely on the relative distances between bodies.
Assuming zero total angular momentum in the center-of-mass frame, the original $3d-$dimensional problem in configuration space $\mathbb{R}^{3d}$ is reduced to a $3-$dimensional space (of relative motion) parametrized by mutual distances. The corresponding six-dimensional Hamiltonian is constructed explicitly in the space of relative motion. A new observation is that
in the \emph{$r-$representation} the Hamiltonian $H_r$ (\ref{Hredzeroam}) at zero potential (the free problem) possesses formally a $\mathbb{Z}_2^{\otimes 3}$-symmetry $r_{ij}\rightarrow -r_{ij}$.
This allows us to code this symmetry by introducing the $\rho_{ij}=r_{ij}^2$ variables and to proceed to
the \emph{$\rho-$representation} obtaining the Hamiltonian ${\cal H}_0$ (\ref{Hredrhozero}).
This Hamiltonian corresponds to the kinetic energy of 3-dimensional solid body. Making the identification of the coefficients of the tensor of inertia in ${\cal H}_{0}$ as the entries of a contravariant metric (cometric), the emerging Hamiltonian describes a 3-dimensional classical particle moving in a curved space with cometric $g^{\mu\,\nu}(\rho)$ (\ref{gmn33-rho}).
Its kinetic energy is a polynomial in coordinate and momentum variables and, unlike the quantum case \cite{TME3-d}, is not accompanied by effective potential. The determinant of cometric Det$\big[g^{\mu\,\nu}(\rho)\big]$ is of definite sign. It vanishes either at the triple-body collision point or on collinear configurations only. This cometric does not become singular at the two-body collision points unlike as it is in the $r-$representation, already noticed in Ref.\cite{Murnaghan} and Ref.\cite{Kampen} for $d=2$ and $d=3$, respectively, as un undesirable property. The free problem in both $r-$ and $\rho-$representations is $\mathcal{S}_3$-permutationally invariant wrt the interchange of body positions and their masses.   \\

The positions of the 3 bodies form a triangle. In the case of three equal masses $m_1=m_2=m_3=1$, the kinetic energy possesses an \emph{accidental} $\mathcal{S}_3$-symmetry wrt the interchange of the edges (intervals of interaction) of this triangle. The full $\mathbb{Z}_2^3 \oplus \mathcal{S}_3^2$ symmetry of the free problem is encoded in the set of generalized coordinates $P$, $S$ and $T$ obtaining the Hamiltonian ${\cal H}_{\rm geo}$ (\ref{Hredgeo}) in the \emph{geometrical-representation}. The two coordinates $P$ and $S$ are called \emph{volume variables}. The variable $P$ corresponds to the sum of squares of sides of the triangle whilst $S$ is the area squared. In $(P,S,T)$-variables, the kinetic energy remains polynomial. It also describes a three-dimensional particle moving in a curved space with a $d$-independent metric $g^{\mu\,\nu}(P,S,T)$. The determinant Det$\big[g^{\mu\,\nu}(P,S,T)\big]$ is again of definite sign. It vanishes when the triangle of interaction is isosceles. The properties and simplicity of geometrical-representation were illustrated in (I) the physically relevant 3-body Newtonian gravity potential ($d=3$) where (\ref{V4g}) describes all four possible 3-body Coulomb/Newton potentials in a unified manner, and (II) in the planar 3-body choreographic motion on algebraic lemniscate ($d=2$) for which the two geometrical variables $P$ and $T$ become \emph{particular constants of motion} and the dynamics is parametrized by a planar elliptic curve $S=\wp (t; P,T)$.\\

The volume variables admit a generalization, $S_m$ (\ref{Sm}) and $P_m$ (\ref{Ptilde}), to systems with arbitrary masses. For potentials that solely depend on these variables, $V=V(P_m,S_m)$, the trajectories in the \emph{volume-representation} are governed by a four-dimensional Hamiltonian ${\cal H}_{\rm vol}$ (\ref{HvolVar}). It describes a two-dimensional particle moving in a curved space. In the volume-representation, the trajectories are mass-independent. In the particular case of a one-variable potential, $V=V(P_m)$, the system is described by a two-dimensional Hamiltonian only. In this representation, a shifted harmonic potential (equivalently, a 3-body mass-dependent anharmonic potential in $r$-variables) was analyzed in detail. The volume-representation implies an effective reduction of the problem beyond separation of variables.  \\

Finally, in the general $n$-body case with $n\geq d-1$ the system can be thought of as a type of "breathing" \emph{polytope of interaction}. The volume-representation can be constructed immediately. At fixed $n$ there exist ($n-1$) volume-variables made out of elements with different dimensionality (edges, faces, cells and so on) of the polytope of interaction. It reveals a surprising link between the theory of regular polytopes and the dynamics of an $n$-body system. This will be presented in a forthcoming paper.

\section{Author's contributions}

All authors contributed equally to this work.

\section{Acknowledgments}

A.M. thanks T.~Fujiwara and E.~Pi\~{n}a for important remarks and personal discussions. R.L. is supported in part by CONACyT grant 237351 (Mexico). A.V.T. thanks Don Saari for a clarification, he is supported in part by the PAPIIT grant {\bf IN113819} (Mexico). W.M. is partially supported by a grant from the Simons Foundation (\# 412351 to Willard Miller,~Jr.).

\section{DATA AVAILABILITY}

Data sharing is not applicable to this article as no new data were created or analyzed in this
study.

%\clearpage

\end{document}